\documentclass[aps,prd,preprint,floatfix,superscriptaddress,nofootinbib,longbibliography,a4paper]{revtex4-1}
\usepackage{graphicx}
\usepackage[T1]{fontenc}
\usepackage[latin9]{inputenc}
\usepackage{amstext}
\usepackage{amsmath,amssymb,array}
\usepackage{babel}
\usepackage{color}
\usepackage{cancel}
\usepackage{hhline}
\usepackage[normalem]{ulem}
\usepackage{hyperref}
\usepackage{empheq}
\usepackage{cases}
\hypersetup{colorlinks=true,
    citecolor=blue,
	linkcolor=blue,
	filecolor=magenta,      
	urlcolor=blue}

\newcommand{\beqa}{\begin{eqnarray}}
\newcommand{\eeqa}{\end{eqnarray}}
\newcommand{\be}{\begin{equation}}
\newcommand{\ee}{\end{equation}}
\newcommand{\ba}{\begin{array}} 
\newcommand{\ea}{\end{array}}

\begin{document}
\title{Electroweak Triplet Scalar Contribution to $SO(10)$ Leptogenesis}

\newcommand{\affUFABC}{Centro de Ci\^encias Naturais e Humanas\;\;\\
	Universidade Federal do ABC, 09.210-170,
	Santo Andr\'e, SP, Brazil}
\newcommand{\affPRL}{Theoretical Physics Division, Physical Research Laboratory, Navarangpura, Ahmedabad-380 009, India}

\author{Chee Sheng Fong}
\email{sheng.fong@ufabc.edu.br}
\affiliation{\affUFABC}

\author{Ketan M. Patel}
\email{ketan.hep@gmail.com}
\affiliation{\affPRL}
	
\begin{abstract}
We show that electroweak triplet scalar can significantly impact baryogenesis via leptogenesis in concrete and predictive $SO(10)$ GUTs, even when light neutrino masses arise predominantly from the type-I seesaw mechanism. This is illustrated within a minimal renormalisable $SO(10)$ model with $\mathbf{10}$ and $\overline{\mathbf{126}}$ scalars in the Yukawa sector and a global Peccei-Quinn-like symmetry. The quark-lepton unification and the flavour structure of the fundamental Yukawa couplings enforce type-I dominance in the light neutrino masses, also suppressing the triplet-induced CP asymmetries in the right-handed neutrino decays. However, the triplet's own decays introduce a new CP-violating source, which can enhance or suppress the total baryon asymmetry. For triplet mass near the right-handed neutrino mass scale, this contribution can dominate, making it essential in assessing the viability of $SO(10)$ leptogenesis scenarios.
\end{abstract}

\maketitle
\flushbottom

\section{Introduction}
Among the ultraviolet completions of the Standard Model (SM) that provide the necessary ingredients to explain the observed matter-antimatter asymmetry via leptogenesis \cite{Fukugita:1986hr}, Grand Unified Theories (GUTs) based on the $SO(10)$ gauge group \cite{Fritzsch:1974nn,GellMann:1980vs} are particularly appealing. In renormalisable $SO(10)$ models, the three generations of singlet fermions, identified as right-handed (RH) neutrinos ($N_i$), naturally emerge as partners of the SM fermions, while an electroweak triplet scalar ${\cal T}$ appears as the counterpart of the SM Higgs doublet. Their presence is thus a generic feature of such frameworks. CP violation is inherently present in the interactions of $N_i$ and ${\cal T}$ through Yukawa couplings, which are generally required to be complex to reproduce the low-energy CP violation observed in the weak interactions of quarks. Below the GUT scale, both $N_i$ and ${\cal T}$ engage in $B-L$ violating processes, and their out-of-equilibrium interactions in the expanding universe can generate a lepton asymmetry, which gets partially converted into a baryon asymmetry via sphaleron processes, see refs.~\cite{Bodeker:2020ghk,DiBari:2021fhs,Xing:2020ald} for the recent reviews.

Leptogenesis under the ``$SO(10)$-inspired'' conditions has been studied extensively~\cite{Buchmuller:1996pa,Nezri:2000pb,Buccella:2001tq,Branco:2002kt,Akhmedov:2003dg,DiBari:2008mp,DiBari:2010ux,Buccella:2012kc,DiBari:2014eya,DiBari:2015oca,DiBari:2017uka,DiBari:2020plh}. These conditions primarily parameterize a relationship between the Dirac neutrino and up-type quark Yukawa couplings. The masses and couplings of the RH neutrinos are then computed from this relation and from the light neutrino masses and mixing, allowing one to estimate the baryon asymmetry. More precise estimations have also been carried out recently in concrete top-down approaches~\cite{Fong:2014gea,Mummidi:2021anm,Patel:2022xxu,Kaladharan:2023zbr,Babu:2024ahk}, in which one does not have to rely on any ad hoc assumptions. Depending on the exact $SO(10)$ model, the masses and couplings of RH neutrinos are determined in terms of a few free parameters of the theory, which can be fixed from the observed spectrum of quarks and leptons in these setups. The estimation of baryon asymmetry in these setups depends on the minimality and predictivity of the models, which are linked to the choice of scalars in the Yukawa sector, the number of light Higgs doublets considered, and the number and phases of the Yukawa couplings. Nevertheless, one obtains a rather precise estimate and can thus check the viability of the underlying model in accounting for the observed matter-antimatter asymmetry. 

However, all these studies consider only the $N_i$-induced lepton asymmetry and neglect the role played by the $SU(2)_L$ triplet scalar ${\cal T}$. One possible reason is the observation that, in the minimal versions of non-supersymmetric $SO(10)$, the neutrino spectrum can be correctly accounted for only if the type-I seesaw contribution dominates in the neutrino masses~\cite{Babu:1992ia,Bajc:2005zf,Joshipura:2011nn,Altarelli:2013aqa,Dueck:2013gca,Meloni:2016rnt,Babu:2016bmy,Ohlsson:2019sja}. Leptogenesis considering the interplay between type-I and type-II seesaw models has also been investigated in refs.~\cite{Chun:2000dr,Joshipura:2001ya,Antusch:2004xy,Antusch:2005tu,AristizabalSierra:2011ab,Pramanick:2022put,Pramanick:2024gvu}, though, to the best of our knowledge, not in the context of a realistic $SO(10)$ GUT. To fill this gap and to assess the contribution of ${\cal T}$ in $SO(10)$ leptogenesis, we carry out this investigation in the context of a minimal model based on ${\bf 10}$ and $\overline{\bf 126}$ scalars in the Yukawa sector and a global $U(1)$ symmetry, which leads to a realistic and predictive setup. One of the main results is that $\cal T$ contribution to leptogenesis is, in general, relevant even if its contribution to light neutrino mass matrix is subdominant.

The organisation of the paper is as follows. In Section~\ref{sec:model}, we review the SO(10) model and quantify the $\cal T$ contribution to the light neutrino mass matrix. In Section~\ref{sec:leptogenesis}, we discuss CP violation in leptogenesis in the presence of both $N_i$ and $\cal T$ and review the lepton-flavour-covariant Boltzmann equations used in the work. In Section~\ref{sec:results}, we present the best fit solution and the determination of baryon asymmetry with and without $\cal T$ contribution. Finally, in Section~\ref{sec:conclusions}, we give a brief summary of the main results of the work.

\section{The SO(10) Model}\label{sec:model}
In an SO(10) GUT model, the SM fermions, together with RH neutrinos, are unified in three generations of $\psi_a$ ($a=1,2,3$) transforming as ${\bf 16}$-dimensional spinorial representations of the unified gauge group. For a minimal realisation, we consider complex ${\bf 10}\sim \Phi$ and $\overline{\bf 126}\sim \overline{\Sigma}$ dimensional scalars in its Yukawa sector. To achieve a greater degree of predictivity, we also consider a global $U(1)$ symmetry under which all $\psi_{a}$ have charge $-1$ while $\Phi$ and $\overline{\Sigma}$ have charge $+2$. The action of $U(1)$ is similar to that of the Peccei-Quinn symmetry \cite{Peccei:1977hh}, and it is shown to be very effective in availing the minimal and realistic Yukawa couplings in the non-supersymmetric versions of $SO(10)$ GUTs \cite{Bajc:2005zf,Joshipura:2011nn}. The most general renormalisable Yukawa interactions under the aforementioned considerations can be parametrised as 
\be \label{LY}
-{\cal L}_Y =  (Y_{10})_{ab}\, \psi_a^T C^{-1} C_5 \Gamma_\mu \psi_b\, \Phi_\mu + \frac{1}{5!} (Y_{\overline{126}})_{ab}\, \psi_a^T C^{-1} C_5 \Gamma_{[\mu} \Gamma_{\nu} \Gamma_{\rho} \Gamma_{\lambda} \Gamma_{\kappa]} \psi_b\, \overline{\Sigma}_{\mu\nu\rho\lambda\kappa}+{\rm h.c.}\,.\ee
Here, $\Gamma_\mu$ ($\mu=1,...,10$) are traceless matrices that act on the spinors of $SO(10)$ and $C_5=\Pi_{i=1}^5 \Gamma_{2i-1}$ \cite{Wilczek:1981iz}. $C$ is the charge conjugation matrix acting on Lorentz spinors. The $3 \times 3$ Yukawa matrices, $Y_{10}$ and $Y_{\overline{126}}$, are in general complex but symmetric in the flavour space as a consequence of fermion statistics and properties of the $\Gamma$ matrices. 

Once $SO(10)$ is broken down to the SM gauge symmetry, $G_{\rm SM}=SU(3)_C \times SU(2)_L \times U(1)_Y$, the interactions in eq.~\eqref{LY} gives rise to the charged and neutral fermion Yukawa interactions as well as to the Majorana masses for the singlet fermions. The scalar $\Phi$ contains a pair of Higgs doublets $h_1\sim (1,2,1/2)$ and $\bar{h}_1\sim(1,2,-1/2)$ under the $G_{\rm SM}$. Similarly, a pair $h_2\sim (1,2,1/2)$ and $\bar{h}_2\sim(1,2,-1/2)$ is contained in $\overline{\Sigma}$. Denoting the quarks and leptons submultiplets residing in $\psi_a$ as $q_a\sim(3,2,1/6)$, $U_a\sim(3,1,2/3)$, $D_a\sim(3,1,-1/3)$, $l_a\sim(1,2,-1/2)$, $E_a\sim(1,1,-1)$ and $N_a\sim(1,1,0)$, the effective Dirac Yukawa interactions take the form \cite{Mummidi:2021anm}:
\beqa \label{LY2}
-{\cal L}_Y & \supset & \overline{q} \left(2\sqrt{2} Y_{10}\,h_1  + 4\sqrt{\frac{2}{3}} Y_{\overline{126}}\,h_2 \right) D\,+\, \overline{\ell} \left(2\sqrt{2} Y_{10}\,h_1  - 4\sqrt{6} Y_{\overline{126}}\,h_2 \right) E\, \nonumber \\
  &+& \overline{q} \left(2\sqrt{2} Y_{10}\,\bar{h}_1  + 4\sqrt{\frac{2}{3}} Y_{\overline{126}}\,\bar{h}_2 \right) U\,+\, \overline{\ell} \left(2\sqrt{2} Y_{10}\,\bar{h}_1  - 4\sqrt{6} Y_{\overline{126}}\,\bar{h}_2 \right) N\,+\,  {\rm h.c.}\,, \eeqa
where we have suppressed the flavour indices for brevity. In addition to the electroweak doublets, $\overline{\Sigma}$ also contains a triplet ${\cal T} \sim (1,3,1)$ and a singlet $\sigma \sim (1,1,0)$ scalars carrying $B-L=2$. Their interactions with the SM fermions, derived from the second term in eq.~\eqref{LY}, are given by
\be \label{LY3}
-{\cal L}_Y \supset \frac{1}{2}\,N^T C^{-1}\,Y_R\,\sigma\,N\,+\,\frac{1}{2}\,\overline{\ell}\,Y_L\, \epsilon{\cal T}^\dagger\,\ell^c\,, \ee
with $Y_R = Y_L = \frac{4}{\sqrt{15}}\,Y_{\overline{126}}$ and 
\begin{eqnarray}
	{\cal T} & = & \left(\begin{array}{cc}
		\frac{1}{\sqrt{2}}{\cal T}^{+} & {\cal T}^{++}\\
		{\cal T}^{0} & -\frac{1}{\sqrt{2}}{\cal T}^{+}
	\end{array}\right)\,.
\end{eqnarray}

Subsequent matching of eq.~\eqref{LY2} with the SM Yukawa interactions depend on the spectrum of the electroweak Higgses and their mixing. To assess this further, consider the most general mass Lagrangian for these doublets, in the basis ${\bf h} = \left(h_1,h_2,\tilde{\bar{h}}_1,\tilde{\bar{h}}_2\right)^T$, which can be parametrized as, 
\be \label{LH}
{\cal L}^{\rm mass}_{\bf h} = -(M^2_{\bf h})_{ij}\, {\bf h}_i^\dagger {\bf h}_j\,.\ee
Here, $\tilde{h}_{1,2} = \epsilon h_{1,2}^*$ and $\tilde{\bar{h}}_{1,2} = \epsilon \bar{h}_{1,2}^*$ with $\epsilon = i\sigma_2$ as the antisymmetric tensor of $SU(2)_L$. The diagonal terms of $M^2_{\bf h}$ can arise from the gauge and $U(1)$ invariant mass terms, for example,
\beqa \label{Mh_diag}
h_1^\dagger h_1  \subset \Phi^* \Phi\,,~~h_2^\dagger h_2  \subset \overline{\Sigma}^*\overline{\Sigma}\,,~~\tilde{\bar{h}}_1^\dagger \tilde{\bar{h}}_1=\bar{h}_1^\dagger \bar{h}_1 \subset \Phi^* \Phi\,,~~\tilde{\bar{h}}_2^\dagger \tilde{\bar{h}}_2=\bar{h}_2^\dagger \bar{h}_2 \subset \overline{\Sigma}^*\overline{\Sigma}\,. \eeqa
The off-diagonal terms, on the other hand, can emerge from gauge symmetry breaking. Some of them would also require $U(1)$ breaking. For example, consider that the scalar sector contains a pair of ${\bf 210}$-dimensional scalars $\Theta^0$, $\Theta$ and a ${\bf 54}$-dimensional $\Omega$. $\Theta^0$ is neutral under $U(1)$ while $\Theta$ and $\Omega$ have charge $+4$. Then six independent off-diagonal terms of $M^2_{\bf h}$ can arise from,
\beqa \label{Mh_offdiag_1}
h_1^\dagger h_2  \subset \Phi^* \overline{\Sigma}\, \langle \Theta^0 \rangle_{1}\,,~h_1^\dagger \tilde{\bar{h}}_1 = (h_1 \cdot \bar{h}_1)^*\subset \Phi^* \Phi^*\, \langle \Omega \rangle_{24}\,,~h_1^\dagger \tilde{\bar{h}}_2 = (h_1 \cdot \bar{h}_2)^*\subset \Phi^* \overline{\Sigma}^*\, \langle \Theta \rangle_{24}\,,\eeqa
and 
\be \label{Mh_offdiag_2}
h_2^\dagger \tilde{\bar{h}}_1 = (h_2 \cdot \bar{h}_1)^*\subset \overline{\Sigma}^* \Phi^*  \langle \Theta \rangle_{1}\,,~h_2^\dagger \tilde{\bar{h}}_2 = (h_2 \cdot \bar{h}_2)^*\subset \overline{\Sigma}^* \overline{\Sigma}^*  \langle \Omega \rangle_{24}\,,~\tilde{\bar{h}}_1^\dagger \tilde{\bar{h}}_2 = (\bar{h}_1^\dagger \bar{h}_2)^*\subset \Phi \overline{\Sigma}^*\langle \Theta^0 \rangle_{24}\,. 
\ee
Here, the subscript in the vacuum expectation values (VEVs) denotes the dimension of $SU(5)$ submultiplet of the given field. The presence of some of these scalars in the complete model is also necessary to break the unified gauge symmetry down to $G_{\rm SM}$ \cite{Bertolini:2009qj}.  

Assuming the most general form for $M^2_{\bf h}$, we postulate that only one of the admixtures of the four Higgs doublets remains light and drives the electroweak symmetry breaking at a scale much below the GUT scale. This is usually achieved through fine-tuning between the scalar potential parameters such that ${\rm Det}(M^2_{\bf h})=0$. Denoting the massless combination of ${\bf h}_i$ as $h$, the latter can be parametrized as
\beqa \label{h}
h &=& c_\theta\, h_1 + s_\theta c_\chi e^{-i \phi_1}\, h_2 + s_\theta s_\chi c_\zeta e^{-i \phi_2}\, \tilde{\bar{h}}_1 + s_\theta s_\chi s_\zeta e^{-i \phi_3}\, \tilde{\bar{h}}_2\,,\eeqa
where $c_\theta = \cos\theta$, $s_\theta = \sin\theta$ and so on. The $h$ is identified with the SM Higgs and substituting the above in eq.~\eqref{LY2}, its effective Yukawa interactions with the fermions can be parameterised as
\beqa \label{eq:L_Yukawa}
-{\cal L}_Y & \supset & \overline{q}\,Y_d\,h\,D\,+\, \overline{q}\,Y_u\,\tilde{h}\,U\, +\, \overline{\ell}\,Y_{e}\,h\, E\,+\, \overline{\ell}\,Y_\nu\,\tilde{h}\,N\, +\, {\rm h.c.}\,, \eeqa
with $\tilde{h} = \epsilon h^*$ and
\beqa \label{Yf}
Y_d &=& H+F, \nonumber \\
Y_e &=& H-3\,F, \nonumber \\
Y_u &=& r\,e^{-i \phi_2}\,\left(H + s\,F\right), \nonumber \\
Y_\nu &=& r\,e^{-i \phi_2}\,\left(H - 3s\, F\right)\,,\eeqa
at the scale of matching, with
\be \label{HFrs}
H = 2\sqrt{2} c_\theta\, Y_{10}\,,~~ F = 4 \sqrt{\frac{2}{3}} s_\theta c_\chi e^{i \phi_1}\, Y_{\overline{126}}\,,~~r=s_\chi c_\zeta \tan\theta\,,~~s=\frac{1}{c_\chi}\, \frac{\tan\zeta}{\tan\theta}\,e^{i(\phi_2-\phi_3-\phi_1)}\,.\ee

Furthermore, the couplings in eq.~\eqref{LY3} can be written as 
\be \label{eq:YL_YR}
Y_R = Y_L = \frac{4}{\sqrt{15}}\,Y_{\overline{126}} \equiv e^{-i \phi_1}\,\frac{F}{q}\,,\ee
with $q = \sqrt{10}\,s_\theta c_\chi$ where we have taken $q>0$ without loss of generality. The light neutrino masses receive contributions from both the type-I and type-II seesaw mechanisms, and it is explicitly given by
\beqa \label{eq:nu_mass}
M_\nu & \simeq &  - \frac{v^2}{\langle \sigma \rangle}\,Y_\nu\,Y_R^{-1}\,Y_\nu^T\,
+ Y_L\,\langle {\cal T}^0\rangle\,.\eeqa
Assuming the most general VEVs, $\langle \sigma \rangle = v_\sigma\, e^{i \phi_\sigma}$ and $\langle {\cal T}^0 \rangle = v_{\cal T}\, e^{i \phi_{\cal T}}$, the effective neutrino mass matrix can be conveniently parametrized as
\beqa \label{eq:nu_mass2}
M_\nu & \simeq &  \frac{v^2}{v_R}\,e^{-i \chi_R}\,\left[-r^2\,(H-3 sF)\,F^{-1}\,(H-3sF)\, +\, r_\nu\,e^{i \chi_\nu}\, F\,\right]\,,\eeqa
where 
\be \label{eq:seesaw_parameters}
v_R = \frac{v_\sigma}{q}\,,~~\chi_R = 2 \phi_2-\phi_1 + \phi_\sigma\,,~~r_\nu = \frac{1}{q^2}\frac{v_\sigma v_{\cal T}}{v^2}\,,~~\chi_\nu = 2 \phi_2 - 2 \phi_1 + \phi_\sigma + \phi_{\cal T}\,.\ee
The parameters $v_R$ and $r_\nu$ are real, and their magnitudes depend on the scalar potential and symmetry-breaking pattern.

The relative strength between the type-I and type-II seesaw contributions to the neutrino masses is parameterised by $r_\nu$.  As it can be seen from the last expression of $M_\nu$, the ratio of the solar to atmospheric squared mass differences and the leptonic mixing parameters do not depend on the overall scale set by $v_R$ and a phase $\chi_R$. However, they do depend on $r_\nu$ and $\chi_\nu$ as the first and second terms possess a different flavour structure.  The condition that the type-II contribution to neutrino mass is subdominant translates to
\be \label{rnu_limit}
r_\nu \ll \frac{r^2}{3}\left|{\rm Tr}(F^{-1} H F^{-1} H)-6s\,{\rm Tr}(F^{-1} H)+27 s^2 \right|.\ee
For $H$ and $F$ of similar order, one finds from above that $r_\nu/r^2 \ll {\cal O}(10)$ indicate type-II seesaw subdominance over the type-I seesaw. 

Remarkably, the quark and lepton spectrum depends on only two complex symmetric matrices $H$ and $F$, dimensionless parameters $r$, $s$, $r_\nu$, $\chi_\nu$ and the scaled $B-L$ breaking scale $v_R$. All these parameters can be more or less determined by matching eqs.~\eqref{Yf} and \eqref{eq:nu_mass2} with the GUT scale extrapolated data of fermion masses and mixing parameters. This in turn allows one to estimate the $B-L$ asymmetries induced in the decays of RH neutrinos and ${\cal T}$ as we discuss in the next section.

\section{Leptogenesis}\label{sec:leptogenesis}
Once the SO(10) gauge symmetry is broken to the SM, the leptonic interactions relevant for leptogenesis can be modeled as
\begin{eqnarray}\label{eq:lag_leptogenesis}
	-{\cal L} & \supset & M_{{\cal T}}^{2}\,{\rm Tr}\left({\cal T}^{\dagger}{\cal T}\right)+\left(\frac{1}{2}f_{\alpha\beta}\,\overline{\ell_{\alpha}^{c}}\epsilon{\cal T}\ell_{\beta}+\frac{1}{2}\mu\, h^{T}\epsilon{\cal T}^{\dagger}h+{\rm h.c.}\right)\nonumber \\
	&  & +\left(\frac{1}{2}(M_R)_{ij}\overline{N_{i}}N_{j}^{c}
	- y_{i\alpha}\overline{N_{i}} h^T \epsilon \ell_{\alpha}  +\left(y_{E}\right)_{\alpha\beta}\overline{E_{\alpha}}h^\dagger \ell_{\beta}  +{\rm h.c.}\right)\,.
\end{eqnarray}
We now use $i,j = 1,2,3$ to denote the flavour indices of the RH neutrinos and $\alpha,\beta = 1,2,3$ the flavour indices of the charged leptons.
Relating these terms with eqs.~\eqref{LY3}, \eqref{eq:L_Yukawa}, \eqref{Yf}, and \eqref{eq:YL_YR}, we get
\begin{eqnarray} \label{eq:yuk_relations}
f &=& \frac{e^{i \phi_1}}{q}\, F^*, \nonumber \\
M_R &=& v_\sigma\,e^{-i \phi_\sigma}\,Y^*_R = v_R\,e^{i (\phi_1 -\phi_\sigma)}\,F^*\,, \nonumber\\
y &=& r e^{i \phi_2} \left(H - 3sF\right)^*, \nonumber\\
y_E &=& (H-3F)^*, \nonumber \\
\mu &=& \frac{2 \langle {\cal T}^0\rangle M_{\cal T}^2}{v^2} = \frac{2 q r_\nu M_{\cal T}^2}{v_R}\,e^{i \phi_{\cal T}}.
\end{eqnarray}
where $q$ is defined below eq.~\eqref{eq:YL_YR} and it can take values in the range $0 < q < \sqrt{10}$. In the last relation, we have matched $2 \langle {\cal T}^0\rangle = \mu v^2/M_{\cal T}^2$ in the type-II seesaw contribution to neutrino mass in eq.~\eqref{eq:nu_mass}.

In the broken phase of the theory, various fields appearing in eq.~\eqref{eq:lag_leptogenesis} can be rephased independently. This allows the overall phases in eq.~\eqref{eq:yuk_relations} to be moved to $\mu$ by the field redefintions:
\be
  N_i \to e^{i(\phi_1 - \phi_\sigma)/2} N_i,\quad
  \ell_\alpha \to  e^{-i\phi_1/2} \ell_\alpha,\quad 
  E_\alpha \to  e^{-i(3\phi_1/2 -  \phi_2 - \phi_\sigma/2)} E_\alpha,\quad 
  h \to  e^{i(\phi_1 - \phi_2 - \phi_\sigma/2)} h,\quad 
\ee
where we obtain the redefined $\mu$ as $\tilde\mu \equiv e^{i\phi}|\mu|$ with the physical CP phase
\be
\phi \equiv \phi_{\cal T} - \phi_\sigma + 2 (\phi_1 - \phi_2). \label{eq:new_CP_phase}
\ee
It becomes clear that CP violation will only depend on $\phi$ if it depends on $\mu$, and this can only happen if it involves $\cal T$. The correctness of eq.~\eqref{eq:new_CP_phase} can be verified explicitly when we determine the contributions to CP violation involving $\cal T$ in the next sections. It will be shown there that these contributions depend only on the combination of couplings $f\mu (y^*)^2$ which in the \emph{mass basis} of $N_i$ gives precisely the overall phase eq.~\eqref{eq:new_CP_phase} as can be checked using eq.~\eqref{eq:yuk_relations}. Consistently, CP violation involving only $N_i$ only involves a combination of couplings $yy^*$ and hence cannot depend on any of the phases in eq.~\eqref{eq:new_CP_phase}.

Assuming $N_i$ are not quasi-degenerate in masses, it is appropriate to work in the mass basis of $N_i$ by writing $M_R = U_M^\dagger \hat{M} U_M^*$ where $U_M$ is a unitary matrix and $\hat M = {\rm diag} (M_1,M_2,M_3)$ with $M_i$ the mass of $N_i$. Without loss of generality, we will fix the ordering $M_1 < M_2 < M_3$. In this basis, we have $\tilde y \equiv U_M y$. 
In the rest of the article, we will work in the $N_i$ mass basis and in the basis in which $\tilde \mu$ contains the overall phase $\phi$. Henceforth, we will also drop the tilde in $\tilde \mu$ and $\tilde y$.

As outlined in the previous section, the parameters $H$, $F$, $r$, $s$, $v_R$ and $r_\nu$ in eq.~\eqref{eq:yuk_relations} can be determined from the fit to fermion masses and mixing while $q$, $M_{\cal T}$ and the phase combination $\phi$ defined in eq.~\eqref{eq:new_CP_phase} cannot be fixed from the fit (only the other phase combination $\chi_\nu$ defined in eq.~\eqref{eq:seesaw_parameters} can be determined from the fit).  Since $q < \sqrt{10}$, if type-I contribution to light neutrino mass dominates $r_\nu \lesssim 10 \,r^2$, one finds $|\mu| \lesssim 2\sqrt{10}\,r^2 M_{\cal T}^2/v_R$. In general, perturbativity imposes a stronger constraint $|\mu| \lesssim M_{\cal T}$.
The remaining parameters, $q$, $M_{\cal T}$ and $\phi$, can be varied to quantify the contribution of ${\cal T}$ in leptogenesis.
\begin{figure}
    \centering
    \includegraphics[width=0.9\linewidth]{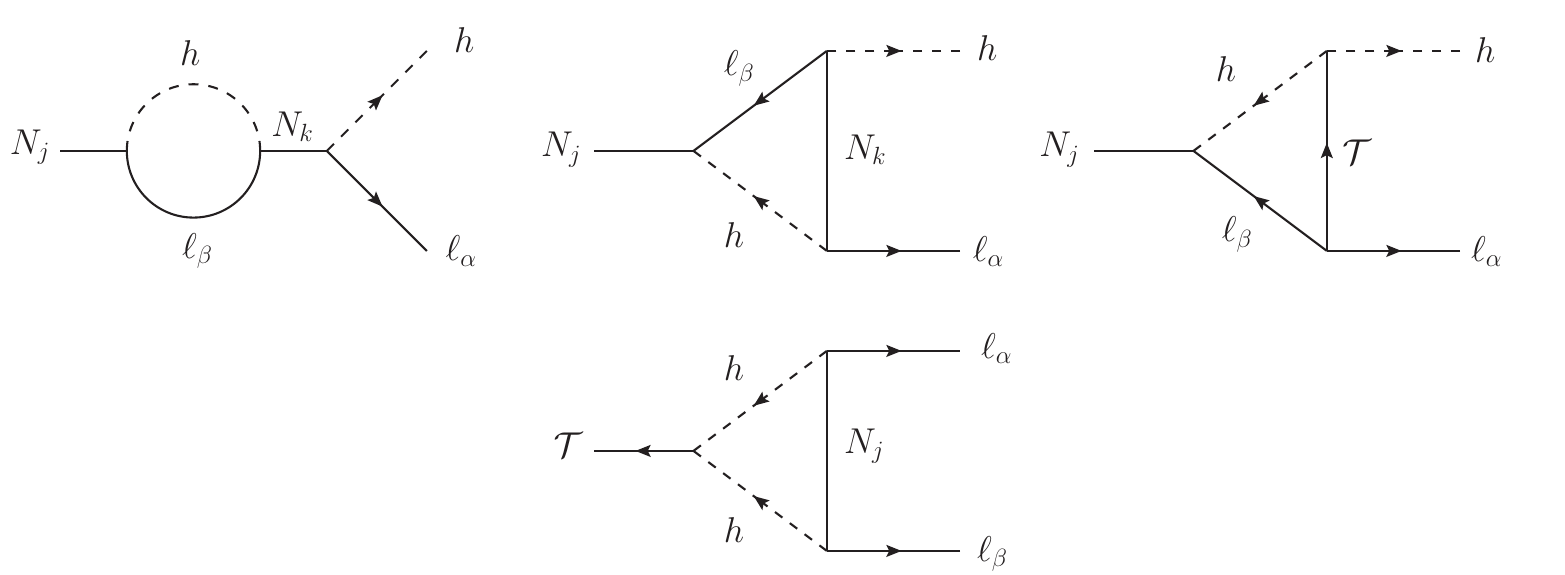}
    \caption{One-loop processes which lead to CP violation in the decays $N_j \to \ell_\alpha h$ (top row) and ${\cal T}^\dagger \to \ell_\alpha \ell_\beta$ (second row).}
    \label{fig:SO10_CP}
\end{figure}

\subsection{Type-I CP violation}
By type-I CP violation, we refer to CP violation in the decays $N_j \to \ell_\alpha h$ (and the CP conjugate processes) which arise from the interference between the tree-level process and one-loop processes as shown in the first row of Fig.~\ref{fig:SO10_CP}.
We define the CP parameter as
\begin{eqnarray}
\epsilon_{j\alpha}^{I} & \equiv & \frac{\Gamma\left(N_j\to\ell_{\alpha}h\right)-\Gamma\left(N_j\to\overline{\ell_{\alpha}}\,h^\dagger\right)}{\Gamma_{N_j}},
\end{eqnarray}
where $\Gamma_{N_j}$ is the total decay width of $N_j$ given by
\be
\Gamma_{N_j} = \frac{(yy^\dagger)_{jj} M_j}{8\pi}.
\ee

The contributions involving $N_{k\neq j}$ as the intermediate state in one-loop processes (the first two diagrams in the first row of Fig.~\ref{fig:SO10_CP}) give~\cite{Covi:1996wh}
\begin{eqnarray}
\epsilon_{j\alpha}^{I,N} & = & -\sum_{k\neq j}\frac{\textrm{Im}\left[\left(yy^{\dagger}\right)_{kj}y_{k\alpha}y_{j\alpha}^{*}\right]}{8\pi\left(yy^{\dagger}\right)_{jj}}g\left(\frac{M_{k}^{2}}{M_{j}^{2}}\right)-\sum_{k\neq j}\frac{\textrm{Im}\left[\left(yy^{\dagger}\right)_{jk}y_{k\alpha}y_{j\alpha}^{*}\right]}{8\pi\left(yy^{\dagger}\right)_{jj}}\frac{M_{j}^{2}}{M_{j}^{2}-M_{k}^{2}},
\label{eq:epIN}
\end{eqnarray}
where
\begin{eqnarray}
	g(x) & = & \sqrt{x}\left[\frac{1}{1-x}+1-\left(1+x\right)\ln\frac{1+x}{x}\right].
\end{eqnarray}

For the contribution of ${\cal T}$ as an intermediate state in the one-loop process (the last diagram in the first row of Fig.~\ref{fig:SO10_CP}), we have~\cite{Hambye:2003ka,Antusch:2004xy}
\begin{eqnarray}
	\epsilon_{j\alpha}^{I,{\cal T}} & = & -\frac{3}{2}\frac{\textrm{Im}\left[\mu\left(fy^\dagger\right)_{\alpha j}y_{j\alpha}^*\right]}{8\pi\left(yy^{\dagger}\right)_{jj}M_{j}}b\left(\frac{M_{{\cal T}}^{2}}{M_{j}^{2}}\right),
    \label{eq:epIT}
\end{eqnarray}
where
\begin{eqnarray}
	b(x) & = & 1-x\ln\frac{1+x}{x}.
\end{eqnarray}
Notice that only combination $\mu f$ appears $\epsilon_{j\alpha}^{I,{\cal T}}$ and from the relations in eq.~\eqref{eq:yuk_relations}, we conclude that the parameter $q$ plays no role. Since $\mu \propto r_\nu$, we have $\epsilon_{j\alpha}^{I,{\cal T}} \propto r_\nu$ as well. Notice that the free phase $\phi$ of $\mu$ allows us to change the overall sign of $\epsilon_{j\alpha}^{I,{\cal T}}$ by taking $\phi \to \phi + \pi$.

From eq.~\eqref{eq:epIN}, the lepton-flavour-covariant CP parameter can be written in term of $3\times 3$ matrix in lepton flavour space as~\cite{Blanchet:2011xq}
\begin{eqnarray}
\left({\cal E}_{j}^{I,N}\right)_{\alpha\beta} & = & \frac{1}{16\pi}\frac{i}{\left(yy^{\dagger}\right)_{jj}}\sum_{k\neq j}\left[\left(yy^{\dagger}\right)_{kj}y_{k\beta}y_{j\alpha}^{*}-\left(yy^{\dagger}\right)_{jk}y_{j\beta}y_{k\alpha}^{*}\right]g\left(\frac{M_{k}^{2}}{M_{j}^{2}}\right)\nonumber \\
 &  & +\frac{1}{16\pi}\frac{i}{\left(yy^{\dagger}\right)_{jj}}\sum_{k\neq j}\left[\left(yy^{\dagger}\right)_{jk}y_{k\beta}y_{j\alpha}^{*}-\left(yy^{\dagger}\right)_{kj}y_{j\beta}y_{k\alpha}^{*}\right]\frac{M_{j}^{2}}{M_{j}^{2}-M_{k}^{2}}.
 \label{eq:epIN_matrix}
\end{eqnarray}
Similarly, from eq.~\eqref{eq:epIT}, we can construct the lepton-flavour-covariant CP parameter matrix as
\begin{eqnarray}
	\left({\cal E}_{j}^{I,{\cal T}}\right)_{\beta\alpha} & = & \frac{3}{2}i\frac{\mu\left(f y^\dagger\right)_{\alpha j}y^*_{j\beta}-\mu^*\left(f y^\dagger\right)_{\beta j}^{*}y_{j\alpha}}{16\pi\left(yy^{\dagger}\right)_{jj}M_{j}}b\left(\frac{M_{{\cal T}}^{2}}{M_{j}^{2}}\right),
     \label{eq:epIT_matrix}
\end{eqnarray}
where we have swapped $\alpha$ and $\beta$ (i.e. the transpose) such that ${\cal E}_{j}^{I,N}$ and ${\cal E}_{j}^{I,{\cal T}}$ will transform the same way under flavour rotations as we will see next. Carrying out rotations in lepton flavour space 
\be
\ell\to U_{\ell}\ell,\quad E\to U_{E}E,
\label{eq:flavour_rotations}
\ee
where $U_\ell$ and $U_E$ are unitary matrices, we have $y\to y U_{\ell}^{\dagger}$, $y_{E}\to U_{E}y_{E}U_{\ell}^{\dagger}$
, $f\to U_{\ell}^{*}fU_{\ell}^{\dagger}$ and the CP parameter matrices transform the same way ${\cal E}_{j}^{I,N}\to U_{\ell}{\cal E}_{j}^{I,N}U_{\ell}^{\dagger}$ and ${\cal E}_{j}^{I,{\cal T}}\to U_{\ell} {\cal E}_{j}^{I,{\cal T}}U_{\ell}^{\dagger}$.

To estimate the relative contributions of ${\cal E}_{j}^{I,N}$ and ${\cal E}_{j}^{I,{\cal T}}$, we will consider their traces, which are independent of lepton flavour basis. The maximum of $|{\rm Tr}\,{\cal E}_{j}^{I,{\cal T}}|$ can be obtained by increasing $\mu$ through increasing $M_{\cal T}$. Considering the limit $M_{\cal T} \gg M_j$, the loop factor $b \to M^2_j/(2M^2_{\cal T})$ and we have
\be
\max |{\rm Tr}\,{\cal E}_{j}^{I,{\cal T}}| \sim \frac{3}{16\pi} \frac{r_\nu M_j |F|}{v_R}, \label{eq:epIT_estimate}
\ee
where $|F|$ denotes the largest absolute value of the matrix entries. 
Taking $M_3 \sim |F| v_R$, the largest $\max|{\rm Tr}\,{\cal E}_{3}^{I,{\cal T}}| \sim r_\nu |F|^2$ can be relevant when $r_\nu$ is sufficiently large. On the other hand, the purely type-I contribution can be estimated as
\be
|{\rm Tr}\,{\cal E}_{j}^{I,N}| \sim \frac{3}{16\pi} |y|^2 r_{ij},
\label{eq:epIN_estimate}
\ee
where $r_{ij} = \min(M_j,M_k)/\max(M_j,M_k)$. The ratio of eqs.~\eqref{eq:epIT_estimate} and \eqref{eq:epIN_estimate} goes like $r_\nu/r^2$
and therefore, $|{\rm Tr}\,{\cal E}_{j}^{I,{\cal T}}|$ is typically suppressed compared to $|{\rm Tr}\,{\cal E}_{j}^{I,N}|$ when the type-I contribution to the light neutrino mass matrix dominates over the type-II contribution.

\subsection{Type-II CP violation}

By type-II CP violation, we refer to CP violation in the decays ${\cal T}^\dagger \to \ell_\alpha \ell_\beta$ (and the CP conjugate processes) which arise from the interference between the tree-level process and one-loop processes which involve $N_j$ as the intermediate states as shown in the second row of Fig.~\ref{fig:SO10_CP}.
The CP parameter for this process is defined as\footnote{The CP violation in the decays ${\cal T} \to hh$ is related to that of ${\cal T}^\dagger \to \ell \ell$ due to unitarity and CPT. In any case, this process is not needed for the calculation since the asymmetry in $h$ will be determined by the in-equilibrium Yukawa interactions.}
\begin{eqnarray}
\epsilon_{\alpha\beta}^{II} & \equiv & \frac{\Gamma\left({\cal T^\dagger}\to\ell_{\alpha}\,\ell_{\beta}\right)-\Gamma\left({\cal T}\to\overline{\ell_{\alpha}}\,\overline{\ell_{\beta}}\right)}{\Gamma_{{\cal T}}+\Gamma_{{\cal T^\dagger}}}\left(1+\delta_{\alpha\beta}\right),
\end{eqnarray}
where tree-level total decay widths of ${\cal T}$ and ${\cal T}^\dagger$ are equal
\be
\Gamma_{\cal T} = \Gamma_{\cal T^\dagger} = \frac{M_{\cal T}}{32\pi}\left[{\rm Tr}(ff^\dagger) + \frac{|\mu|^2}{M_{\cal T}^2}\right],
\ee
and for later use, we will define the decay branching ratios to lepton and Higgs final states, respectively, as
\begin{eqnarray}
B_{\ell} & = & \frac{{\rm Tr}\left(ff^{\dagger}\right)}{{\rm Tr}\left(ff^{\dagger}\right)+\left|\mu\right|^{2}/M_{{\cal T}}^{2}},\quad
B_{h} =\frac{\left|\mu\right|^{2}/M_{{\cal T}}^{2}}{{\rm Tr}\left(ff^{\dagger}\right)+\left|\mu\right|^{2}/M_{{\cal T}}^{2}}. \label{eq:BR}
\end{eqnarray}

We obtain~\cite{Hambye:2003ka,AristizabalSierra:2014nzr}
\begin{eqnarray}
\epsilon_{\alpha\beta}^{II} & = & 
\frac{1}{4\pi}\sum_{j}M_{j}\frac{\textrm{Im}\left(\mu f_{\alpha\beta}y_{j\alpha}^{*}y_{j\beta}^{*}\right)}{M_{{\cal T}}^{2}\textrm{Tr}\left(ff^{\dagger}\right)+\left|\mu\right|^{2}}\ln\left(1+\frac{M_{{\cal T}}^{2}}{M_{j}^{2}}\right).
\label{eq:epII_alpha_beta}
\end{eqnarray}
The source term for $\alpha$ flavour is obtained by summing over $\beta$ 
\begin{eqnarray}
\sum_{\beta}\epsilon_{\alpha\beta}^{II} & = & \frac{1}{4\pi}\sum_{j}M_{j}\frac{\textrm{Im}\left[\mu\left(fy^{\dagger}\right)_{\alpha j}y_{j\alpha}^{*}\right]}{M_{{\cal T}}^{2}\textrm{Tr}\left(ff^{\dagger}\right)+\left|\mu\right|^{2}}\ln\left(1+\frac{M_{{\cal T}}^{2}}{M_{j}^{2}}\right).
\label{eq:epII_alpha}
\end{eqnarray}
In the numerator, only the combination $\mu f$ appears and hence $q$ and $r_\nu$ only affect the decay branching ratios of $\cal T$ through $\mu$ in eq.~\eqref{eq:yuk_relations}. For smaller (larger) $q$ and $r_\nu$, $B_\ell$ is enhanced (suppressed) while to $B_h$ is suppressed (enhanced). For instance, to obtain equal branching ratios to both final states, we have
\be \label{eq:qq}
 q^2 = \frac{\sqrt{{\rm Tr}(FF^\dagger)}v_R}{2r_\nu M_{\cal T}}.
\ee
Just like $\epsilon_{j\alpha}^{I,{\cal T}}$, the overall sign of $\epsilon_{j\alpha}^{II}$ can be changed by taking $\phi \to \phi + \pi$.

From eq.~\eqref{eq:epII_alpha}, we can make the CP parameter lepton-flavour-covariant by writing it as~\cite{Lavignac:2015gpa}
\begin{eqnarray}
{\cal E}_{\beta\alpha}^{II} & = & -\frac{i}{8\pi}\sum_{j}M_{j}\frac{\mu\left(fy^{\dagger}\right)_{\alpha j}y_{j\beta}^{*}-\mu^{*}\left(fy^{\dagger}\right)_{\beta j}^{*}y_{j\alpha}}{M_{{\cal T}}^{2}\textrm{Tr}\left(ff^{\dagger}\right)+\left|\mu\right|^{2}}\ln\left(1+\frac{M_{{\cal T}}^{2}}{M_{j}^{2}}\right).
\end{eqnarray}
We have again swapped $\alpha$ and $\beta$ such that under flavour rotations specified in eq.~\eqref{eq:flavour_rotations}, we have
${\cal E}^{II}\to U_\ell{\cal E}^{II}U_\ell^\dagger$ which transforms just like ${\cal E}^{I,N}$ and ${\cal E}^{I,{\cal T}}$.

Considering the lepton-flavour-basis-independent trace of ${\cal E}_{j}^{II}$, we have
\be
\max |{\rm Tr}\,{\cal E}^{II}| \sim \frac{|y|^2}{4\pi} \sqrt{B_\ell B_h}\frac{M_3}{M_{\cal T}}\ln\left(1+\frac{M_{{\cal T}}^{2}}{M_{3}^{2}}\right).\label{eq:epII_estimate}
\ee
For small $M_{\cal T}/M_3$, $B_\ell \gg B_h$ and $|{\rm Tr}\,{\cal E}^{II}|$ is suppressed by
\be
\delta \equiv \frac{2 q r_\nu M_{\cal T}^2}{|f|v_R M_3} = \frac{2q^2 r_\nu M_{\cal T}^2}{|F|v_R M_3},\label{eq:epII_suppression}
\ee
while for large $M_{\cal T}/M_3$, $B_h \gg B_\ell$ and the suppression factor is $1/\delta$. The maximum of $|{\rm Tr}\,{\cal E}^{II}|$ occurs for $\delta \sim 1$ or around $M_{\cal T} \sim M_3/\sqrt{2q^2 r_\nu}$ which is \emph{not} equivalent to the condition of equal branching ratios eq.~\eqref{eq:qq} and is comparable in magnitude to the largest $|{\rm Tr}\,{\cal E}_j^{I,N}|$.

\subsection{Boltzmann equations} \label{subsec:BE}
Let us first define the number asymmetry of particle $i$ as
\begin{eqnarray}
Y_{\Delta i} & \equiv Y_{i}-Y_{\bar{i}}= & \frac{n_{i}}{s}-\frac{n_{\bar{i}}}{s},
\end{eqnarray}
where $n_i$ and $n_{\bar i}$ are number density of $i$ and its antiparticle $\bar i$, respectively, and $s=2\pi^{2}g_{\star}T^{3}/45$ is the cosmic entropy density where $g_{\star}$ is the effective relativistic degrees of freedom. Then, the baryon asymmetry is given by 
\begin{eqnarray}
Y_B & = & \sum_{i}q_{i}^{B}Y_{\Delta i},
\end{eqnarray}
where $q_{i}^{B}$ is the baryon number carried by $i$. For the observed value, we will take the Planck measurement of baryon number~\cite{Planck:2018vyg} as the reference value: $Y_B^{\rm CMB} = (8.70 \pm 0.06) \pm 10^{-11}$.

For the generation of baryon asymmetry through leptogenesis, we will utilize the lepton-flavour-covariant formalism developed in ref.~\cite{Fong:2021xmi} where we will need to describe the asymmetries in both $\ell$ and $E$ through the $3\times 3$ matrices of asymmetry in number density in flavour space $(Y_{\Delta \ell})_{\alpha\beta}$ and $(Y_{\Delta E})_{\alpha\beta}$. Under flavour rotations of eq.~\eqref{eq:flavour_rotations}, they transform as $U_\ell Y_{\Delta \ell} U_\ell^\dagger$ and $U_E Y_{\Delta E} U_E^\dagger$ and the total asymmetry in $\ell$ and $E$ are given respectively by ${\rm Tr}(Y_{\Delta \ell})$ and ${\rm Tr}(Y_{\Delta E})$, which are clearly flavour basis independent. Defining the following charge combination
\begin{eqnarray}
Y_{\widetilde{\Delta}} & \equiv & \frac{1}{3}Y_{B}I_{3\times3}-Y_{\Delta\ell},\label{eq:tildeDelta_matrix}
\end{eqnarray}
they satisfy the following flavour-covariant Boltzmann equations due to interactions mediated by charged lepton Yukawa
\begin{eqnarray}
s{\cal H}z\frac{dY_{\widetilde{\Delta}}}{dz} & = & \frac{\gamma_{E}}{2Y^{{\rm nor}}}\left\{ y_{E}^{\dagger}y_{E},\frac{Y_{\Delta\ell}}{g_{\ell}\zeta_{\ell}}\right\} -\frac{\gamma_{E}}{Y^{{\rm nor}}}y_{E}^{\dagger}y_{E}\frac{Y_{\Delta h}}{g_{h}\zeta_{h}}-\frac{\gamma_{E}}{Y^{{\rm nor}}}y_{E}^{\dagger}\frac{Y_{\Delta E}}{g_{E}\zeta_{E}}y_{E},\label{eq:BE_YtildeDelta} \\
s{\cal H}z\frac{dY_{\Delta E}}{dz} & = & -\frac{\gamma_{E}}{2Y^{{\rm nor}}}\left\{ y_{E}y_{E}^{\dagger},\frac{Y_{\Delta E}}{g_{E}\zeta_{E}}\right\} -\frac{\gamma_{E}}{Y^{{\rm nor}}}y_{E}y_{E}^{\dagger}\frac{Y_{\Delta h}}{g_{h}\zeta_{h}}+\frac{\gamma_{E}}{Y^{{\rm nor}}}y_{E}\frac{Y_{\Delta\ell}}{g_{\ell}\zeta_{\ell}}y_{E}^{\dagger},\label{eq:BE_YE}
\end{eqnarray}
where the corresponding reaction density $\gamma_{E}$ was determined in refs.~\cite{Garbrecht:2013bia,Garbrecht:2014kda} to be $\gamma_{E}\approx5\times10^{-3}T^4/6$, the anticommutator is $\left\{ A,B\right\} \equiv AB+BA$,
$z\equiv M_{\textrm{ref}}/T$ with $M_{\textrm{ref}}$ an arbitrary
mass scale, and ${\cal H}=1.66\sqrt{g_{\star}}T^{2}/M_{\textrm{Pl}}$
is the radiation-dominated Hubble rate with $M_{\rm Pl} = 1.22 \times 10^{19}$ GeV. We have also defined $Y^{{\rm nor}}\equiv 15/(8\pi^{2}g_{\star})$
with $g_{i}$ is the gauge degrees of freedom of $i$ ($g_{\ell}=g_{h}=2$
and $g_{E}=1$) and
\begin{eqnarray}
\zeta_{i} & \equiv & \frac{6}{\pi^{2}}\int_{m_{i}/T}^{\infty}dx \, x\sqrt{x^{2}-m_{i}^{2}/T^{2}}\frac{e^{x}}{\left(e^{x}+\xi_{i}\right)^{2}},
\end{eqnarray}
with $m_{i}$ the mass of $i$ and $\xi_{i}=1(-1)$ if $i$ is a fermion (boson).
Since we are considering a temperature range much above the electroweak symmetry breaking scale, we will set $m_{\ell}=m_{E}=m_{h}=0$, which gives
$\zeta_{\ell}=\zeta_{E}=\zeta_{H}/2=1$. 

To realise leptogenesis, we will add to eq.~\eqref{eq:BE_YtildeDelta} and/or eq.~\eqref{eq:BE_YE} the corresponding source terms. Since both processes $N_j \to \ell_\alpha h$ and ${\cal T}^\dagger \to \ell_\alpha \ell_\beta$ violate $\tilde \Delta$ charge, we will add the corresponding source terms to eq.~\eqref{eq:BE_YtildeDelta} 
\begin{eqnarray}
S^{{\rm I}} & \equiv & -\sum_{j}\left({\cal E}^{I,N}_{j} + {\cal E}^{I,{\cal T}}_{j} \right)\gamma_{N_{j}}\left(\frac{Y_{N_{j}}}{Y_{N_{j}}^{{\rm eq}}}-1\right) + \frac{1}{2}\sum_{j}\frac{\gamma_{N_{j}}}{Y^{{\rm nor}}}\left(\frac{1}{2}\left\{ P_{j},\frac{Y_{\Delta\ell}}{g_{\ell}\zeta_{\ell}}\right\} +P_{j}\frac{Y_{\Delta h}}{g_{h}\zeta_{h}}\right), \\
S^{{\rm II}} & \equiv & -{\cal E}^{II}\gamma_{\cal T}\left(\frac{Y_{\Sigma{\cal T}}}{Y_{\Sigma{\cal T}}^{{\rm eq}}}-1\right) \nonumber \\
& &+ \frac{2\gamma_{\cal T}}{{\rm Tr}\left(ff^{\dagger}\right)+\left|\mu\right|^{2}/M_{{\cal T}}^{2}}
\left[f^*f^{T}\frac{Y_{\Delta{\cal T}}}{Y_{{\cal T}}^{{\rm eq}}}+\frac{2f^* Y_{\Delta\ell}^{T}f^{T}+f^*f^{T}Y_{\Delta\ell}+Y_{\Delta\ell}f^*f^{T}}{4Y^{{\rm nor}}g_{\ell}\zeta_{\ell}}\right],
\end{eqnarray}
supplemented by the Boltzmann equations to describe the evolutions of $N_j$, ${\cal T}$ and ${\cal T}^\dagger$
\begin{eqnarray}
s{\cal H}z\frac{dY_{N_{j}}}{dz}&=& -\gamma_{N_{j}}\left(\frac{Y_{N_{j}}}{Y_{N_{j}}^{{\rm eq}}}-1\right),\\
s{\cal H}z\frac{dY_{\Sigma{\cal T}}}{dz} & = & -\gamma_{\cal T}\left(\frac{Y_{\Sigma{\cal T}}}{Y_{\Sigma{\cal T}}^{{\rm eq}}}-1\right)-2\gamma_{A}\left(\frac{Y_{\Sigma{\cal T}}^{2}}{Y_{\Sigma{\cal T}}^{{\rm eq},2}}-1\right),\\
s{\cal H}z\frac{dY_{\Delta{\cal T}}}{dz} & = & -\gamma_{\cal T}\left(\frac{Y_{\Delta{\cal T}}}{Y_{\Sigma{\cal T}}^{{\rm eq}}}+B_{\ell}\frac{{\rm Tr}\left(ff^{\dagger}Y_{\Delta\ell}\right)}{{\rm Tr}\left(ff^{\dagger}\right)Y^{{\rm nor}}g_{\ell}\zeta_{\ell}}-B_{h}\frac{Y_{\Delta h}}{Y^{{\rm nor}}g_{h}\zeta_{h}}\right),
\end{eqnarray}
where $Y_{\Sigma {\cal T}} \equiv Y_{\cal T} + Y_{{\cal T}^\dagger}$ and  $Y_{\Delta {\cal T}} \equiv Y_{\cal T} - Y_{{\cal T}^\dagger}$. For $N_j$ and $\cal T$, we have taken their equilibrium abundances to be that of Maxwell-Boltzmann  $Y_{N_{i}}^{{\rm eq}}=\frac{45}{2\pi^{4}g_{\star}}\frac{M_{i}^{2}}{T^{2}}{\cal K}_{2}\left(\frac{M_{i}}{T}\right)$ and $Y_{\Sigma{\cal T}}^{{\rm eq}}=Y_{{\cal T}}^{{\rm eq}}+Y_{{\cal T}^{\dagger}}^{{\rm eq}}=\frac{135}{2\pi^{4}g_{\star}}z^{2}{\cal K}_{2}\left(z\right)$ with ${\cal K}_{n}\left(x\right)$ the modified Bessel function of the second kind of order $n$ and the reference mass scale is set to $M_{\cal T}$ such that $z = M_{\cal T}/T$.
Assuming Maxwell-Boltzmann distributions in the reaction densities, we have included the decay and inverse decay processes for $N_j$ and $\cal T$
\begin{eqnarray}
\gamma_{N_{j}} & = & sY_{N_{j}}^{{\rm eq}}\Gamma_{N_{j}}\frac{{\cal K}_{1}\left(M_{j}/T\right)}{{\cal K}_{2}\left(M_{j}/T\right)}, \\
\gamma_{\cal T} & = & sY_{\Sigma {\cal T}}^{{\rm eq}}\Gamma_{{\cal T}}\frac{{\cal K}_{1}\left(z\right)}{{\cal K}_{2}\left(z\right)},
\end{eqnarray}
and scatterings of ${\cal T}{\cal T}^\dagger$ with the SM fields through  $SU(2)_L \times U(1)_Y$ gauge interactions
\begin{eqnarray}
\gamma_{A} & = & \frac{M_{{\cal T}}^{4}}{64\pi^{4}z}\int_{4}^{\infty}dx\sqrt{x}\,{\cal K}_{1}\left(z\sqrt{x}\right)\hat{\sigma}_{A}\left(x\right),
\end{eqnarray}
where the reduced cross section is given by~\cite{Hambye:2005tk}
\begin{eqnarray}
\hat{\sigma}_{A}\left(x\right) & = & \frac{1}{16\pi x^{2}}\left\{ \sqrt{x}\sqrt{x-4}\left[96g_{2}^{2}g_{Y}^{2}\left(x+4\right)+g_{Y}^{4}\left(65x-68\right)+2g_{2}^{4}\left(172+65x\right)\right]\right.\nonumber \\
 &  & \!\!\!\!\! \left.-96\left[4g_{2}^{2}g_{Y}^{2}\left(x-2\right)+g_{Y}^{4}\left(x-2\right)+4g_{2}^{4}\left(x-1\right)\right]\ln\left(\frac{\sqrt{x-4}\sqrt{x}+x}{2}-1\right)\right\} .
\end{eqnarray}

To have a closed set of equations, the asymmetries in $\ell$ and $h$ can be written as~\cite{Fong:2021xmi}
\begin{eqnarray}
Y_{\Delta\ell} & = & \frac{2}{15}c_{B}I_{3\times3}{\rm Tr}Y_{\widetilde{\Delta}}-Y_{\widetilde{\Delta}},\\\label{eq:Yellmatrix_Ycharges}
Y_{\Delta h} & = & -c_{H}\left({\rm Tr}Y_{\widetilde{\Delta}}-2{\rm Tr}Y_{\Delta E}+2Y_{\Delta{\cal T}}\right),\label{eq:YH_Ycharges_HiggsTriplets}
\end{eqnarray}
where the temperature-dependent coefficients $c_B$ and $c_H$, which depend on which SM interactions are in equilibrium, can be parametrised as
\begin{eqnarray}
c_{B}\left(T\right) & = & 1-e^{-\frac{T_{B}}{T}},\label{eq:cB}\\
c_{H}\left(T\right) & = & \begin{cases}
1 & T>T_{t}\\
\frac{2}{3} & T_{u}<T<T_{t}\\
\frac{14}{23} & T_{u-b}<T<T_{u}\\
\frac{2}{5} & T_{u-c}<T<T_{u-b}.\\
\frac{4}{13} & T_{B_{3}-B_{2}}<T<T_{u-c}\\
\frac{3}{10} & T_{u-s}<T<T_{B_{3}-B_{2}}\\
\frac{1}{4} & T_{u-d}<T<T_{u-s}\\
\frac{2}{11} & T<T_{u-d}
\end{cases},\label{eq:cH}
\end{eqnarray}
with $T_B = 2.3 \times 10^{12}$ GeV, $T_t = 10^{15}$ GeV, $T_u = 2\times 10^{13}$ GeV, $T_{u-b} = 3\times 10^{11}$ GeV, $T_{u-c} = 2\times 10^{10}$ GeV, $T_{B_3-B_2} = 9\times 10^{8}$ GeV, $T_{u-s} = 3\times 10^{8}$ GeV and $T_{u-d} = 2\times 10^{6}$ GeV. 

After leptogenesis, the total $B-L$ charge given by
\be
 Y_{B-L} = {\rm Tr}\,Y_{\widetilde{\Delta}}-{\rm Tr}\,Y_{\Delta E},
 \label{eq:Y_B-L}
\ee
is conserved. The lattice calculation of ref.~\cite{DOnofrio:2014rug} indicates that the electroweak sphaleron processes freeze out at 132 GeV after the electroweak symmetry breaking at 160 GeV, and therefore the final baryon asymmetry can be written as
\begin{eqnarray}
Y_{B} & = & c_{sp}(T=132\,{\rm GeV})\, Y_{B-L},
\end{eqnarray}
where taking into account of finite mass of the top quark $m_t$, we have~\cite{Fong:2015vna}
\be
 c_{sp}(T) = \frac{6(5+\zeta_t)}{97+14\zeta_t}.
\ee
Using $m_t = 173$ GeV, we obtain $c_{sp} = 0.315$, which is the value we will adopt.

\section{Results}\label{sec:results}
Our aim in this section is to estimate the values of various parameters appearing in eqs. \eqref{Yf} and \eqref{eq:nu_mass2} through detailed numerical fits to fermion masses and mixing observables, and subsequently use them to quantify the contributions to the lepton asymmetries. 

\subsection{Fitting the quark and lepton spectra}
Utilising the freedom to choose the basis for $\psi_a$ in eq.~\eqref{LY}, the matrix $H$ can be made diagonal and real. It is straightforward to see that the matrix $F$ remains symmetric in this basis. The observable quantities, such as the fermion masses and mixing, do not depend on the overall phases $\phi_2$ and $\chi_R$ in eqs.~\eqref{Yf} and \eqref{eq:nu_mass2}. This leaves a total of 21 real parameters, i.e. 3 in $H$, 12 in $F$, $r$, Re($s$), Im($s$), $r_\nu$, $\chi_\nu$ and $v_R$, which determine the 22 low-energy observables in principle. Among the latter, the three CP phases in the lepton sector and an absolute neutrino mass scale have not been measured directly. We use the remaining 18 observables to estimate the values of the aforementioned parameters using the usual $\chi^2$-optimisation technique discussed at length in \cite{Mummidi:2021anm}.
 
We first evolve the charged fermion Yukawa couplings and quark mixing parameters from their low-energy values at the top-quark pole mass \cite{Mummidi:2021anm} to the GUT scale, $M_{\rm GUT} = 2 \times 10^{16}$ GeV, using 1-loop renormalization group evolution (RGE) equations of the SM. We consider normal or inverted ordering in the neutrino masses and evaluate both cases separately. Moreover, we include only the ratio $\Delta m_{21}^2/\Delta m_{31}^2$ in the definition of $\chi^2$, allowing the latter to bea  function of only the dimensionless parameters. The parameter $v_R$ is determined subsequently by comparing the theoretically estimated value of $\Delta m_{31}^2$ with its experimental one. As the RGE effects on the ratio $\Delta m_{21}^2/\Delta m_{31}^2$ and leptonic mixing observables are insignificant \cite{Antusch:2003kp,Chankowski:2001mx,Ohlsson:2013xva}, we adopt the low-energy values of these observables from \cite{Esteban:2024eli} while matching eqs.~\eqref{Yf} and \eqref{eq:nu_mass2} at $M_{\rm GUT}$. 

The experimental inputs (evolved to GUT scale) used in the $\chi^2$ function are listed as $O_{\rm exp}$ in Table \ref{tab:bestfit}. Instead of considering the RGE extrapolated standard deviations for these observables in the definition of $\chi^2$, we assign a $30\%$ uncertainty in the light quark masses and $10\%$ in all the other observables. This conservative range is intended to accommodate potential higher-order corrections, threshold effects and influence of possible intermediate scales. Such effects are model-dependent and their explicit evaluations require specification of a complete model beyond the Yukawa sector considered in the present work. Nonetheless, we expect that the chosen error margins absorb these uncertainties without substantially impacting our results.

\begin{figure}
    \centering
    \includegraphics[width=0.46\linewidth]{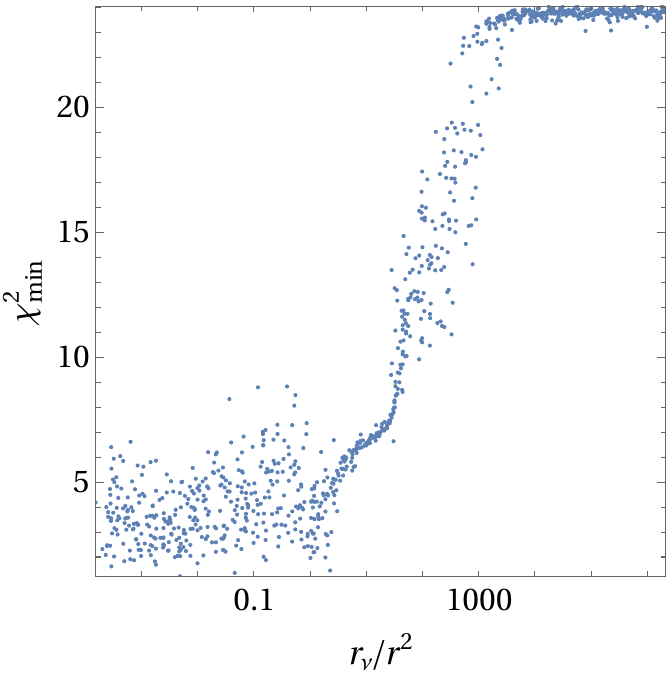}\hspace*{0.4cm}
    \includegraphics[width=0.47\linewidth]{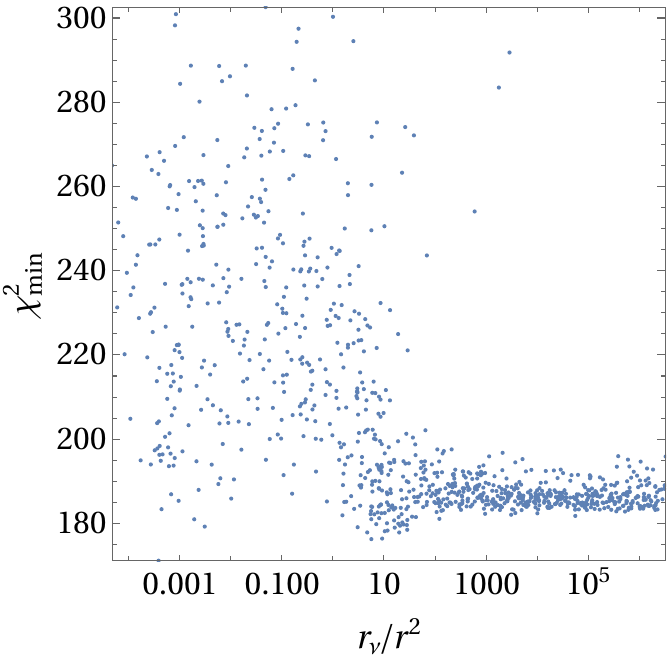}
    \caption{Minimised $\chi^2$ as a function of $r_\nu/r^2$ for normal (left panel) and inverted (right panel) ordering in the light neutrino masses. Values of $\chi^2_{\rm min} < 9$ disfavour $r_\nu/r^2 > 20$ in the case of normal ordering, thereby ruling out a dominant type-II seesaw contribution to the neutrino masses. The inverted ordering case is entirely disfavoured.} 
    \label{fig:chs_rnu}
\end{figure}

To quantify the relative contributions of type-I and type-II seesaw mechanisms to the light neutrino masses, we fix the parameter $r_\nu$ to a particular value which can be chosen positive without loss of generality, and optimise the $\chi^2$ function with respect to the other parameters. The values of minimised $\chi^2$, i.e. $\chi^2_{\rm min}$, for different $r_\nu$ obtained in this way are displayed in Fig.~\ref{fig:chs_rnu} for the normal and inverted ordering in the left and right panels, respectively.
As it can be seen, a relatively good $\chi^2_{\rm min}$ is obtained only if $r_\nu/r^2 \lesssim 10$ in case of normal ordering. In comparison with eq.~\eqref{rnu_limit}, the latter implies dominance of type-I contribution to the neutrino masses. This reaffirms the earlier results \cite{Joshipura:2011nn,Dueck:2013gca} indicating the failure of type II seesaw mechanism as the sole source of neutrino masses in the minimal non-supersymmetric $SO(10)$ frameworks. Moreover, the result in Fig.~\ref{fig:chs_rnu} puts a quantitative upper limit on $r_\nu$, which translates to an upper limit on $\mu$ through eq.~\eqref{eq:yuk_relations}. It can also be seen that the inverted ordering case is entirely disfavoured as it fails to provide acceptable $\chi^2_{\rm min}$. The normal ordering in neutrino masses is preferred in the present setup due to the inherent quark-lepton unification.

For practical purposes, we consider $\chi^2_{\rm min} < 9$ a reasonably good fit as none of the observables deviates more than $3\sigma$ from $O_{\rm exp}$. The best fit solution from all the points in Fig.~\ref{fig:chs_rnu} we get, corresponds to $\chi^2_{\rm min} = 1.2$ and the resulting values of the observables are given in Table \ref{tab:bestfit}. We further list the predictions for the leptonic CP phases, mass of the lightest neutrino $m_{\nu_1}$, effective mass of the electron-neutrino $m_{\beta} = \left(\sum_i m_{\nu_i}^2 |U_{ei}|^2\right)^{1/2}$, the effective Majorana mass of electron-neutrino $m_{\beta \beta} = |\sum_i m_{\nu_i} U_{ei}^2|$ and the masses of RH neutrinos offered by the solution in Table~\ref{tab:predictions}. For the definition of CP phases, we use the convention used in \cite{Patel:2022xxu}.
We find no preference for any particular value or range for $\delta_{\rm PMNS}$ among the solutions with $\chi^2_{\rm min} < 9$. This is a typical feature of this kind of models, also seen earlier in refs.~\cite{Joshipura:2011nn,Feruglio:2015iua,Babu:2018tfi,Mummidi:2021anm}, and it is due to the presence of a sizable number of complex parameters. The specific predictions for the leptonic CP phases can be obtained if additional restrictions on the phases of the fundamental Yukawa couplings are imposed, see \cite{Patel:2022xxu} for example. For the observables involving the absolute mass scale of neutrinos, we find the predicted ranges: $m_{\nu_1} \in[3,9]\,{\rm meV}$, $m_{\beta} \in [9,13]\,{\rm meV}$ and $m_{\beta \beta} \in [0.1,0.6]\,{\rm meV}$.
\begin{table}[ht!]
	\begin{center} 
		\begin{math} 
			\begin{tabular}{cccc}
				\hline
				\hline 
				~~~Observables~~~ & ~~~~~~~~$O_{\rm th}$~~~~~~~~  & ~~~~~~~~$O_{\rm exp}$~~~~~~~~  & ~~~Pull~~~ \\
				\hline
				$y_u$  & $2.96 \times 10^{-6}$ & $2.92\times 10^{-6}$ & $\sim 0$ \\
				$y_c$   & $1.49\times10^{-3}$& $1.47\times 10^{-3}$& $0.1$\\
				$y_t$   & $0.438 $ & $0. 444$& $-0.1$\\
				$y_d$  & $ 4.58 \times 10^{-6} $ & $6.42\times10^{-6}$  & $-1.0$ \\
				$y_s$   & $1.17 \times10^{-4}$ & $1.28\times10^{-4}$ & $-0.3$ \\
				$y_b$   & $5.76 \times10^{-3}$ & $5.86 \times10^{-3}$& $-0.2$ \\
				$y_e$   & $2.77 \times 10^{-6}$ & $2.76 \times 10^{-6}$ & $0.1$ \\
				$y_{\mu}$  &$5.84 \times10^{-4}$ & $5.75\times10^{-4}$ & $0.2$\\
				$y_{\tau}$   & $9.93 \times10^{-3}$ & $9.72 \times10^{-3}$& $0.2$\\
				$\Delta m^2_{\text{21}} [{\rm eV}^2]$ & $7.53\times10^{-5}$& $7.49\times10^{-5}$ & $0.1$\\
				$\Delta m^2_{\text{31}} [{\rm eV}^2]$ & $2.534\times10^{-3}$& $2.534\times10^{-3}$& $0$\\
				$|V_{us}|$ & 0.2325 & 0.2304 & $0.1$\\
				$|V_{cb}|$ & 0.0478 & 0.0484 & $-0.1$ \\
				$|V_{ub}|$ & 0.0042 & 0.0043 & $-0.1$ \\
				$\sin\delta_{\rm CKM}$ & 0.904 & 0.910 & $-0.1$  \\
				$\sin^2 \theta _{12}$ & 0.313 & 0.307 & $0.2$\\
				$\sin^2 \theta _{23}$ & 0.551   & 0.561 & $-0.2$ \\
				$\sin^2 \theta _{13}$ & 0.02190 & 0.02195  & $\sim 0$ \\	  	
				\hline
				\hline 
			\end{tabular}
		\end{math}
	\end{center}
	\caption{Best fit results at the GUT scale obtained for an example solution corresponding to $\chi_{\rm min}^2 = 1.2$.} 
	\label{tab:bestfit} 
\end{table}
\begin{table}[ht!]
	\begin{center} 
		\begin{math} 
			\begin{tabular}{cccc}
				\hline
				\hline
				\multicolumn{4}{c}{Predictions} \\
				\hline
				$\sin\delta_{\rm PMNS}$ & $-0.56$ & \quad $M_{1}$ [GeV]  & $7.14 \times 10^{9}$ \\
				$\sin\eta_1$ & $-0.25$   & \quad $M_{2}$ [GeV]  & $2.79 \times 10^{11}$ \\
				$\sin\eta_2$ & $-0.83$  & \quad $M_{3}$ [GeV]  & $ 1.43 \times 10^{12}$ \\
				$m_{\nu _1} $[meV] &  $6.2$  & \quad $m_{\beta \beta}$ [meV]  &  $0.44$ \\
				$ m_{\beta } $[meV]  & $10.8$  & & \\			
				\hline
				\hline 
			\end{tabular}
		\end{math}
	\end{center}
	\caption{Predictions obtained for an example solution corresponding to $\chi_{\rm min}^2 = 1.2$.} 
	\label{tab:predictions} 
\end{table}

The fitted values of various parameters obtained for the best fit solution are:
\beqa \label{eq:H_F_values}
H &=& \left(
\begin{array}{ccc}
 0.000257 & 0. & 0. \\
 0. & -0.056806 & 0. \\
 0. & 0. & 6.73755 \\
\end{array}
\right)\times 10^{-3}\,, \nonumber \\
F &=& \left(
\begin{array}{ccc}
 -0.001214+0.002247 i & 0.005611\, -0.011005 i & -0.091735+0.054459 i \\
 0.005611\, -0.011005 i & 0.14708\, +0.06613 i & 0.325945\, -0.157876 i \\
 -0.091735+0.054459 i & 0.325945\, -0.157876 i & -1.00934+0.29367 i \\
\end{array}
\right)\times 10^{-3}\,,\nonumber \\ \eeqa
and
\be \label{eq:other_values}
r=-67.8665\,,~s=0.2984 - 0.0274 i\,,~r_\nu=22.3872\,,~\chi_\nu=1.1016\,,~v_R=1.2273 \times 10^{15}\,{\rm GeV}\,.\ee
This corresponds to $r_\nu/r^2=0.0049$, indicating the strong type-I dominance in the neutrino masses for the present solution.  Using the values above, we have
\be
\frac{|\mu|}{M_{\cal T}} = 0.018 \left(\frac{q}{0.5}\right)\left(\frac{M_{\cal T}}{10^{12}\,{\rm GeV}}\right),
\ee
where the perturbativity bound $|\mu|/M_{\cal T} \lesssim 1$ is satisfied.

\subsection{Estimation of baryon asymmetry}

Using the best fit values of eq.~\eqref{eq:H_F_values} and \eqref{eq:other_values}, we can determine the baryon asymmetry produced from leptogenesis by solving the Boltzmann equations described in Section \ref{subsec:BE}. Before doing so, we will first study the magnitudes of CP parameters to understand when triplet scalar contributions will be relevant. Since a good fit requires type-I dominance in light neutrino mass contribution, we can take $r_\nu$ as an \emph{upper bound} while $q$, $\phi$ and $M_{\tau}$ are the remaining free parameters which determine the triplet scalar contribution to leptogenesis.
For convenience of the readers, we will reproduce $y$ and $y_E$ in the mass basis of $N_j$ here
\beqa \label{eq:y_yE_values}
y &=& \left(
\begin{array}{ccc}
 0.000312 + 0.000133i & 0.000432+0.000076i & 0.030454-0.007283i \\
 0.000880+0.000816i & -0.016814-0.001377i & 0.123818+0.036829i \\
 0.003090-0.005643i & -0.010841+0.016423i & -0.043428-0.498744i\\
\end{array}
\right)\,, \nonumber \\
y_E &=& \left(
\!\!\!\begin{array}{ccc}
 0.000390+0.000674 i & -0.001683 - 0.003301i & 0.027521+0.016338i \\
 -0.001683 -0.003301 i & -0.049805 +0.019839 i & -0.097783 -0.047363 i \\
 0.027521 + 0.016338 i & -0.097783 -0.047363 i & 0.976556 + 0.088101 i \\
\end{array}
\right)\times 10^{-2}\,. \nonumber \\ \eeqa

\begin{figure}
    \centering
    \includegraphics[width=0.48\linewidth]{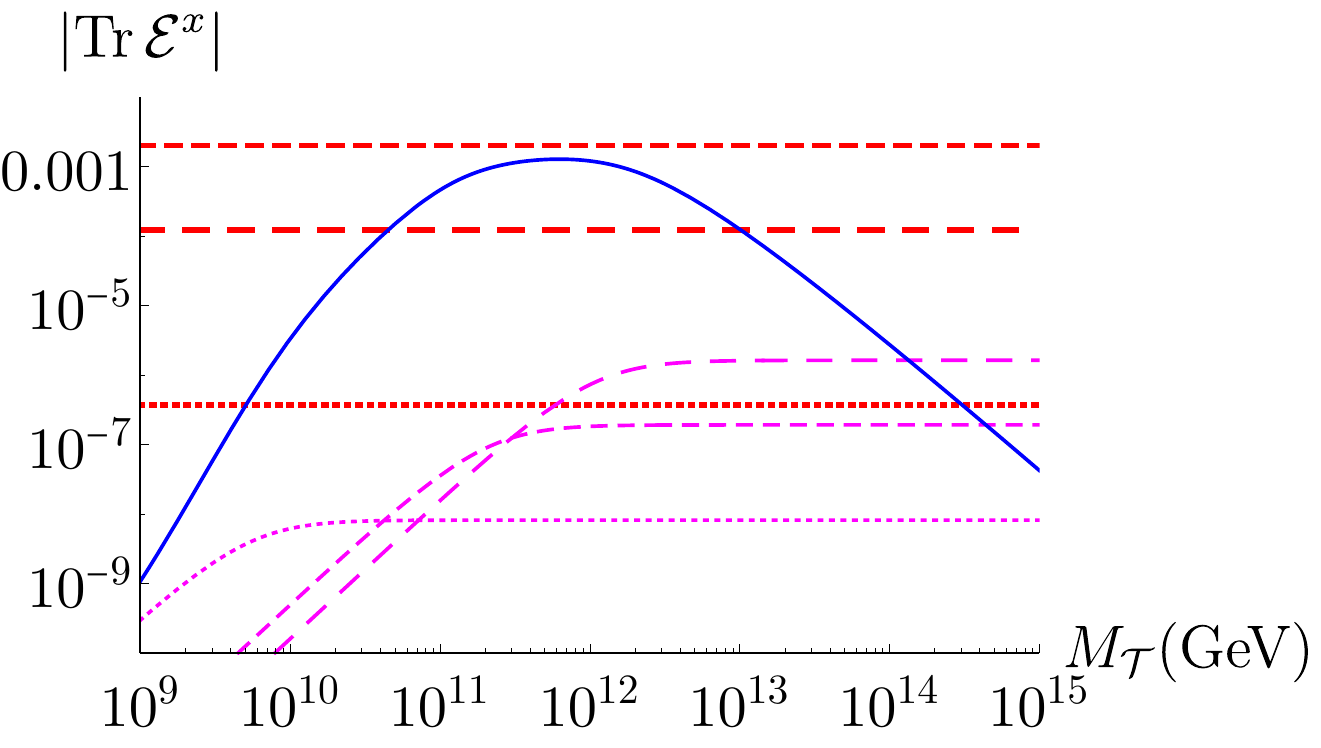}
    \includegraphics[width=0.48\linewidth]{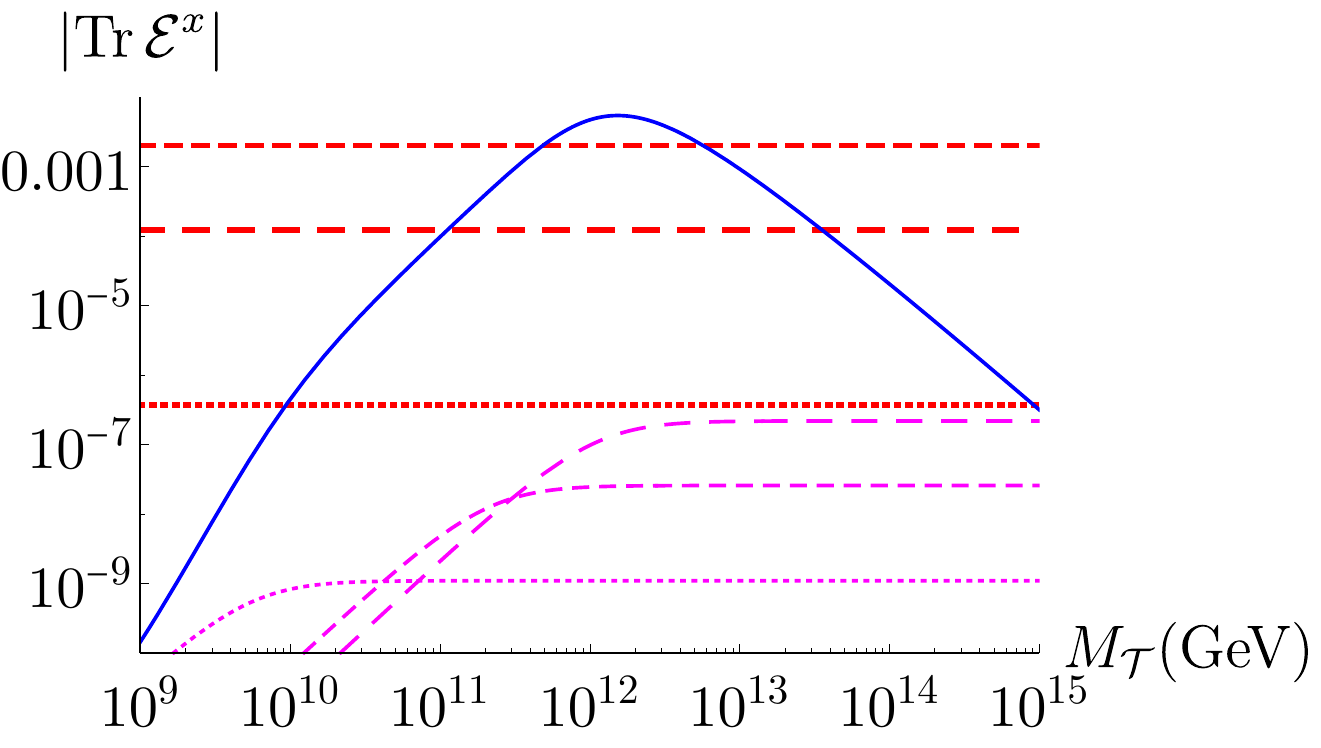}
    \caption{$|{\rm Tr}\,{\cal E}^x|$ as a function of $M_{\cal T}$ for $x=I,N$ (thick red), $x=I,{\cal T}$ (thin magenta) and $x=II$ (solid blue) fixing $q = 0.5$ and $\phi = 4$ for a benchmark best-fit point with $r_\nu = 22.3872$ (left plot) and $r_\nu = 3$ (right plot). The dotted, dashed and long-dashed curves are CP parameters from decays of $N_1$, $N_2$ and $N_3$, respectively. While CP parameters for $N_1$ and triplet contribution to decays of $N_j$ are subdominant, the contribution from decay of $\cal T$ can be important for $M_{\cal T} \sim 10^{12}$ GeV.}
    \label{fig:TrCP_comparison}
\end{figure}

\begin{figure}
    \centering
    \includegraphics[width=0.48\linewidth]{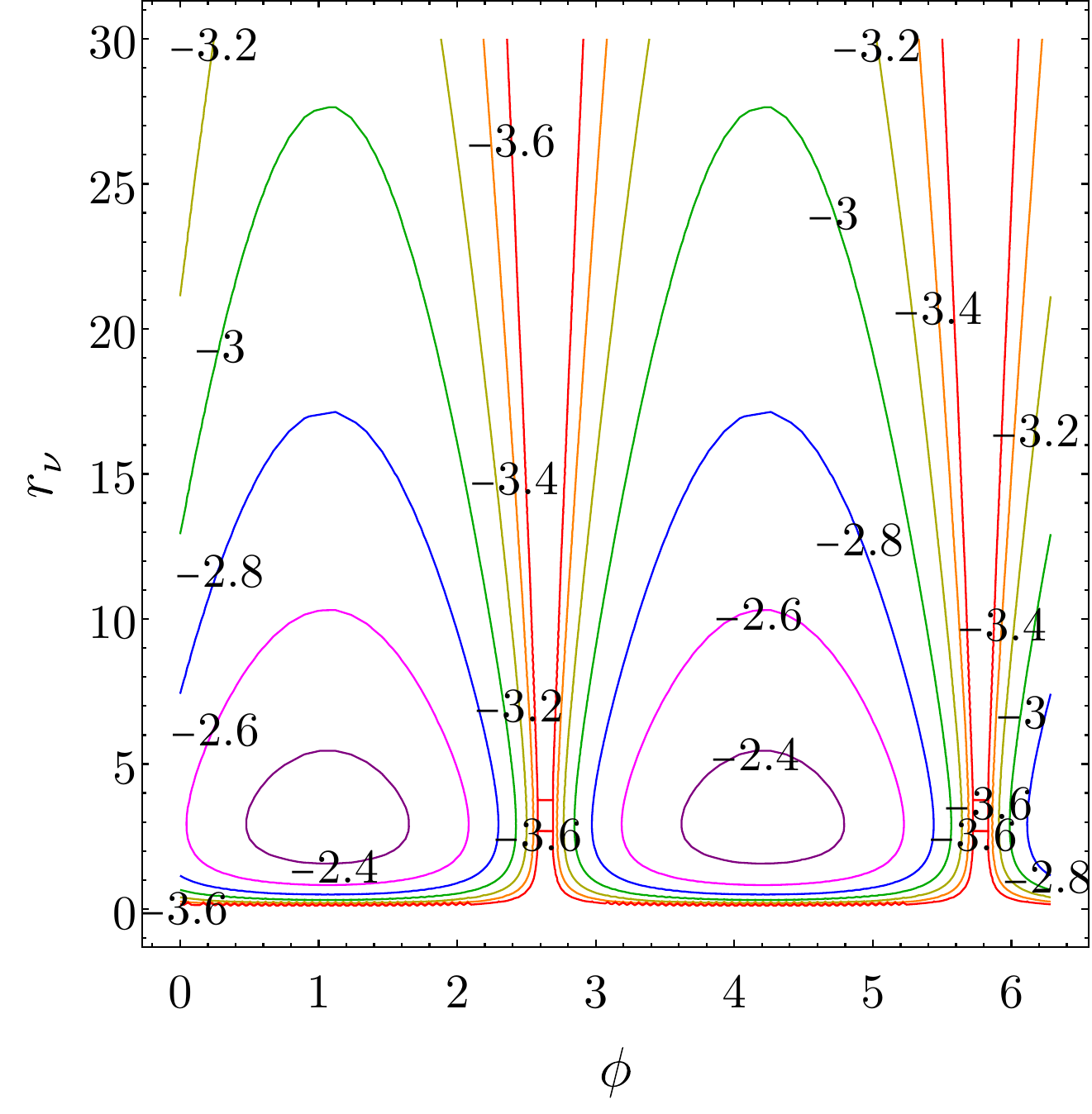}\hspace{0.5cm}
    \includegraphics[width=0.48\linewidth]{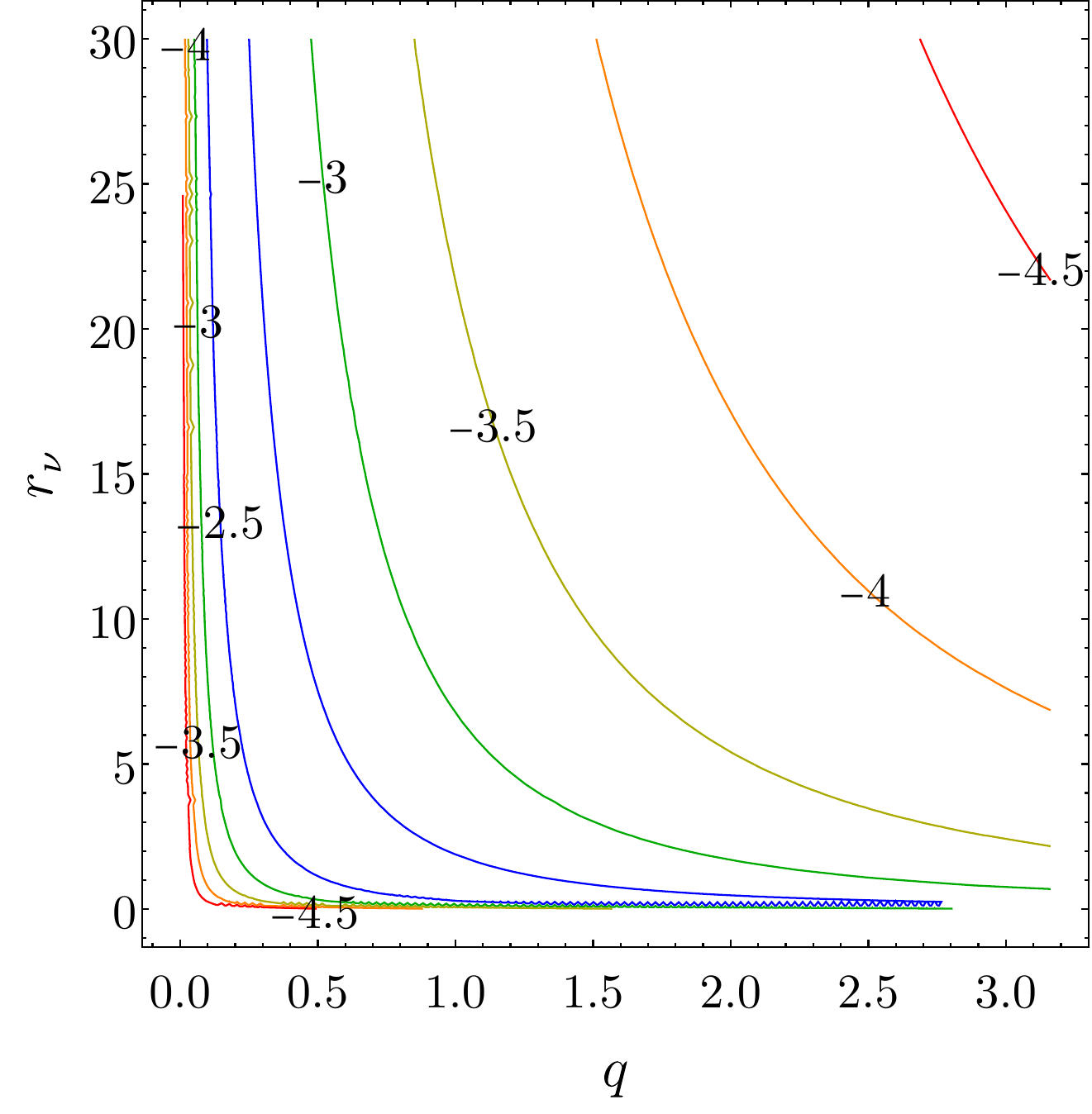}
    \caption{Contours of $\log|{\rm Tr}\,{\cal E}^{II}|$ as a function of $r_\nu$ and $\phi$ fixing $q = 0.5$ (left) and as a function of $r_\nu$ and $q$ fixing $\phi = 4$ (right). In both cases, we have fixed $M_{\cal T} = 10^{12}$ GeV.}
    \label{fig:q_phi_rnu_dependence}
\end{figure}

In Fig.~\ref{fig:TrCP_comparison}, we plot $|{\rm Tr}\,{\cal E}_j^{I,N}|$, $|{\rm Tr}\,{\cal E}_j^{I,{\cal T}}|$ and $|{\rm Tr}\,{\cal E}^{II}|$ as a function of $M_{\cal T}$ using the best fit solution and further fixing $q=0.5$ and $\phi=4$. As references, we have plot $|{\rm Tr}\,{\cal E}_j^{I,N}|$ for $j=1,2,3$ (dotted, dashed and long-dashed red horizontal lines) which are independent of $q$, $\phi$, $r_\nu$ and $M_{\tau}$. 
For  $|{\rm Tr}\,{\cal E}_j^{I,{\cal T}}|$ with $j = 1,2,3$ (dotted, dashed and long-dashed magenta curves), we see that they asymptote to some maximum values as $M_{\cal T}$ increases as explained around eq.~\eqref{eq:epIT_estimate}. Their overall magnitude is proportional to $r_\nu$: in the left plot, we use the best fit value $r_\nu = 22.3872$ while on the right we take a smaller value $r_\nu = 3$. $|{\rm Tr}\,{\cal E}_j^{I,{\cal T}}|$ remains subdominant in this model since the best fit point gives $|y| \gg |F|$. Finally, we see that $|{\rm Tr}\,{\cal E}^{II}|$ have an expected maximum around $M_{\cal T} \sim M_3 \sim 10^{12}$ GeV with magnitude comparable to the largest $|{\rm Tr}\,{\cal E}_j^{I,N}|$ as explained around eq.~\eqref{eq:epII_estimate}.

Focusing on $|{\rm Tr}\,{\cal E}^{II}|$, we study its dependence on $\phi$, $q$, and $r_\nu$ in Fig.~\ref{fig:q_phi_rnu_dependence}. In the left plot, we see that for a fixed $q$, $|{\rm Tr}\,{\cal E}^{II}|$ first increases with $r_\nu$ ($B_\ell \gg B_h$) and then decreases with $r_\nu$ ($B_h \ll B_\ell$) while $\phi$ determines its overall sign as expected for a phase. In particular, one can change the overall sign by taking $\phi \to \phi + \pi$. The dependency on $r_\nu$ can be understood from the dependency on $\delta$ defined in eq.~\eqref{eq:epII_suppression} and similar behaviour will hold by fixing $r_\nu$ while varying $q$. In the right plot, fixing $\phi = 4$, we show the dependency of $|{\rm Tr}\,{\cal E}^{II}|$ on $q$ and $r_\nu$ where the contours satisfy $r_\nu \propto 1/q^2$ as can again be understood from eq.~\eqref{eq:epII_suppression}.

\begin{figure}
    \centering
    \includegraphics[width=0.48\linewidth]{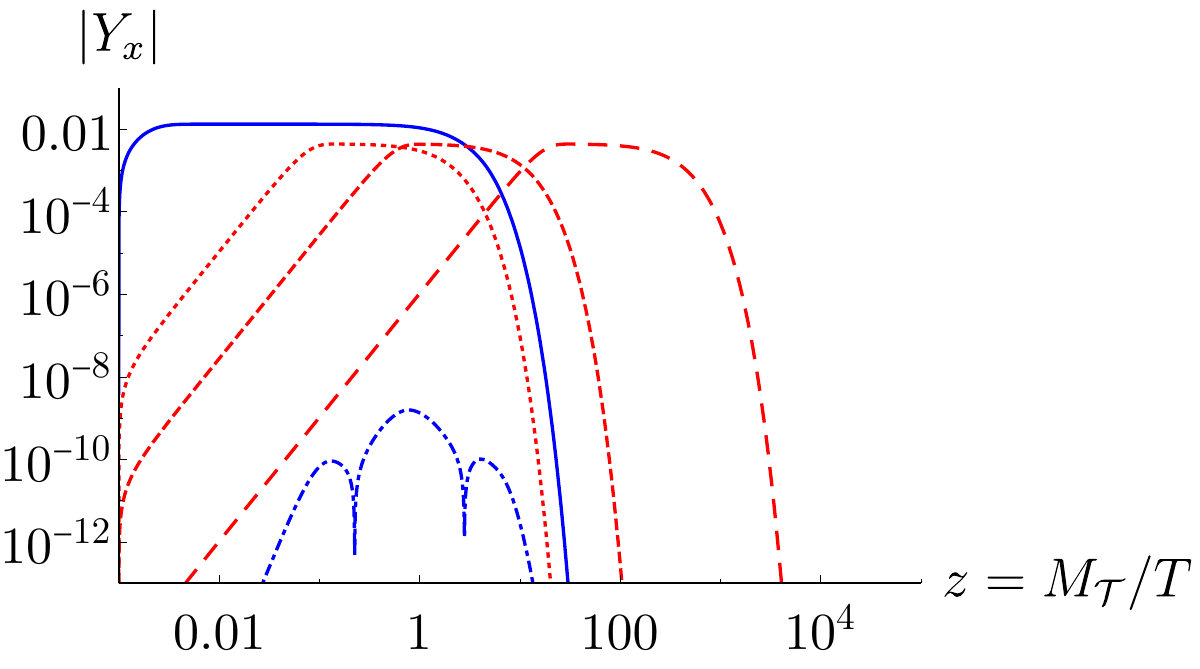}
    \includegraphics[width=0.48\linewidth]{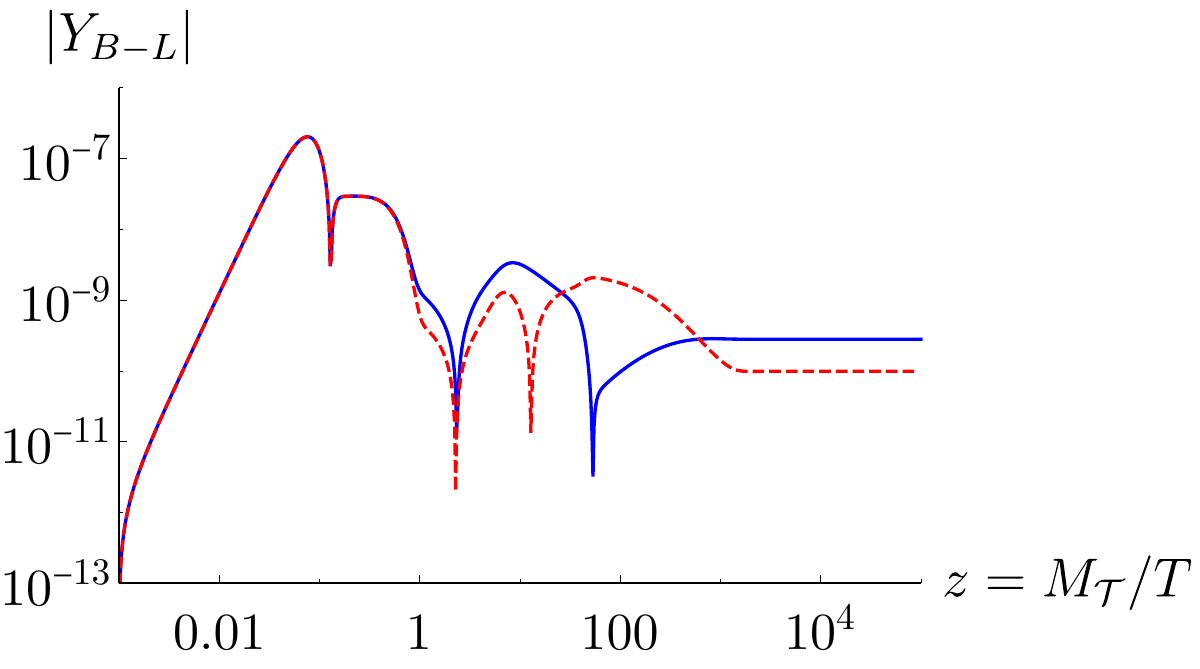}
    \caption{In the left plot, we have the evolutions of $Y_{N_1}$ (long dashed red curve), $Y_{N_2}$ (dashed red curve), $Y_{N_3}$ (dotted red curve), $Y_{\Sigma {\cal T}}$ (solid blue curve) and $|Y_{\Delta {\cal T}}|$ (dot-dashed blue curve) as a function of $z = M_{\cal T}/T$. In the right plot, we have the evolutions of $|Y_{B-L}| = |{\rm Tr}\,Y_{\widetilde{\Delta}}-{\rm Tr}\,Y_{\Delta E}|$ as a function of $z = M_{\cal T}/T$ for purely $N_j$ contribution to leptogenesis (red blue curve) and with the additional contribution from $\cal T$ (solid blue curve) fixing $M_{\cal T} = 10^{12}$ GeV, $q = 0.5$ and $\phi = 4.76$ such that $Y_B = 0.315Y_{B-L}=8.8\times 10^{-11}$. The final $Y_{B-L}$ values in both scenarios are independent of initial abundances of $\cal T$ and $N_j$ since they are quickly thermalised in the early phase.} 
    \label{fig:YB_comparison}
\end{figure}

Finally, in Fig.~\ref{fig:YB_comparison}, we show as a function of $z = M_{\cal T}/T$, the evolutions of $Y_{N_1}$ (long dashed red curve), $Y_{N_2}$ (dashed red curve), $Y_{N_3}$ (dotted red curve), $Y_{\Sigma {\cal T}}$ (solid blue curve) and $|Y_{\Delta {\cal T}}|$ (dot-dashed blue curve) in the left plot and the evolutions of $|Y_{B-L}| = |{\rm Tr}\,Y_{\widetilde{\Delta}}-{\rm Tr}\,Y_{\Delta E}|$ in the right plot. The red-blue curve shows the evolution of $Y_{B-L}$ from purely $N_j$ contributions to leptogenesis, which give a final $Y_B = 3.1 \times 10^{-11}$, which is about a factor of three smaller than the observed value. Adding the contribution of $\cal T$ by choosing $M_{\cal T} = 10^{12}$ GeV, $q = 0.5$ and $\phi = 4.76$, the evolution of $Y_{B-L}$ is shown as the solid blue curve which gives a final $Y_B = 8.8 \times 10^{-11}$. Comparing the solid red and dashed blue curves, one can see the contribution from decays of $\cal T$ from $z \gtrsim 10$. In both cases, $N_1$ plays a negligible role in producing an asymmetry (due to suppressed CP violation) but gives an important washout effect which further reduces the asymmetry generated up to $z \sim 10^3$. We have checked that the final values of $Y_{B-L}$ in both scenarios are independent of initial abundances of $\cal T$ and $N_j$, the former due to fast gauge interactions while the latter due to large $|y|$, which thermalises the $N_j$ before their decays.

\section{Conclusions}\label{sec:conclusions}
We have quantified the contribution of $SU(2)_L$ triplet scalar $\cal T$ to leptogenesis in a realistic and predictive SO(10) GUT in this work. The model contains only ${\bf 10}$ and $\overline{\bf 126}$ scalars in the Yukawa sector, supplemented by a global $U(1)$ symmetry. Only one linear combination of the electroweak doublet Higgs arising from ${\bf 10}$ and $\overline{\bf 126}$ is assumed to be light, which drives the electroweak symmetry breaking. The $\overline{\bf 126}$ also contains an SM singlet and $SU(2)_L$ triplet scalars whose VEVs break $B-L$ and give Majorana masses to heavy and light neutrinos, respectively. The couplings of all these scalars with the quarks and leptons are determined only in terms of two fundamental Yukawa coupling matrices due to inherent quark-lepton unification. As a result, the couplings of electroweak triplet $\cal T$ to the SM fields are more or less completely determined up to a mixing parameter $0<q<\sqrt{10}$, a scalar potential parameter $|\mu| \lesssim M_{\cal T} $ and its phase $\phi$. 

In the context of leptogenesis, our main observations are:
\begin{itemize}
    \item A good fit in this model requires type-I seesaw contribution to dominate the contribution to light neutrino mass matrix with hierarchical right-handed neutrino masses $M_1 \sim 10^{10}$ GeV, $M_2 \sim 10^{11}$ GeV and $M_3 \sim 10^{12}$ GeV. For type-I CP violation, dominant CP violation comes from decays of $N_2$ while $N_1$ is relevant for washout of the generated asymmetry. 
    
    \item For type-II CP violation, one-loop contribution of $\cal T$ to the decays of right-handed neutrinos $N_i$ remains subdominant while the CP violation from decays of $\cal T$ dominate when its mass $M_{\cal T}$ is of the same order as the mass of the heaviest right-handed neutrino $M_3$. In SO(10) models where the type-II seesaw contribution to the light neutrino mass matrix is important, one-loop contribution of $\cal T$ to $N_i$ decays will be important.
    
    \item Besides $M_{\cal T}$, the CP violation involving $\cal T$ depends on order of $q$, $|\mu|/M_{\cal T}$ and $\phi$. While $|\mu|$ is bounded from above by the type-II seesaw contribution to the light neutrino mass matrix, the free phase $\phi$ allows both positive and negative contributions to the final baryon asymmetry. In particular, one should not despair in the case where purely type-I leptogenesis is not sufficient to explain the baryon asymmetry (as is the case for our best fit solution), since $\cal T$ contribution can make leptogenesis viable again.
\end{itemize}

In summary, over a range of parameters, both ${\cal T}$ and $N_i$ can contribute significantly to leptogenesis in realistic $SO(10)$ models, despite the subdominance of the type-II seesaw contribution to neutrino masses. This subdominance arises from the relatively weak coupling of ${\cal T}$ to leptons compared to the Dirac neutrino Yukawa couplings, which are typically of the same order as the up-type quark Yukawa couplings. With regard to predictions, we do not find a strong preference for the Dirac CP phase in the leptonic sector or for a particular octant of $\theta_{23}$. The effective Majorana mass relevant for neutrinoless double beta decay is predicted to lie in the range $0.1$-$0.6$ meV, well below the current upper limit of $36$-$156$ meV from the KamLAND-Zen experiment~\cite{KamLAND-Zen:2022tow}. The sum of light neutrino masses is predicted to be in the range $62$-$70$ meV, which remains consistent with current cosmological bounds~\cite{Shao:2024mag}.

\begin{acknowledgments}
    We would like to acknowledge support from the ICTP through the Associates Programme, where this work was initiated during our visit to ICTP-Trieste in 2023. We would also like to thank Diego Aristizabal for discussion on type-II CP violation. CSF acknowledges the support by Funda\c{c}\~ao de Amparo \`a Pesquisa do Estado de S\~ao Paulo (FAPESP) Contract No. 2019/11197-6 and Conselho Nacional de Ci\^encia e Tecnologia (CNPq) under Contract No. 304917/2023-0. The work of KMP is supported by the Department of Space (DOS), and partially under the MATRICS project (MTR/2021/000049) from the Science \& Engineering Research Board (SERB), Department of Science and Technology (DST), Government of India.
\end{acknowledgments}

\bibliography{references}

\begin{thebibliography}{61}%
\makeatletter
\providecommand \@ifxundefined [1]{%
 \@ifx{#1\undefined}
}%
\providecommand \@ifnum [1]{%
 \ifnum #1\expandafter \@firstoftwo
 \else \expandafter \@secondoftwo
 \fi
}%
\providecommand \@ifx [1]{%
 \ifx #1\expandafter \@firstoftwo
 \else \expandafter \@secondoftwo
 \fi
}%
\providecommand \natexlab [1]{#1}%
\providecommand \enquote  [1]{``#1''}%
\providecommand \bibnamefont  [1]{#1}%
\providecommand \bibfnamefont [1]{#1}%
\providecommand \citenamefont [1]{#1}%
\providecommand \href@noop [0]{\@secondoftwo}%
\providecommand \href [0]{\begingroup \@sanitize@url \@href}%
\providecommand \@href[1]{\@@startlink{#1}\@@href}%
\providecommand \@@href[1]{\endgroup#1\@@endlink}%
\providecommand \@sanitize@url [0]{\catcode `\\12\catcode `\$12\catcode
  `\&12\catcode `\#12\catcode `\^12\catcode `\_12\catcode `\%12\relax}%
\providecommand \@@startlink[1]{}%
\providecommand \@@endlink[0]{}%
\providecommand \url  [0]{\begingroup\@sanitize@url \@url }%
\providecommand \@url [1]{\endgroup\@href {#1}{\urlprefix }}%
\providecommand \urlprefix  [0]{URL }%
\providecommand \Eprint [0]{\href }%
\providecommand \doibase [0]{http://dx.doi.org/}%
\providecommand \selectlanguage [0]{\@gobble}%
\providecommand \bibinfo  [0]{\@secondoftwo}%
\providecommand \bibfield  [0]{\@secondoftwo}%
\providecommand \translation [1]{[#1]}%
\providecommand \BibitemOpen [0]{}%
\providecommand \bibitemStop [0]{}%
\providecommand \bibitemNoStop [0]{.\EOS\space}%
\providecommand \EOS [0]{\spacefactor3000\relax}%
\providecommand \BibitemShut  [1]{\csname bibitem#1\endcsname}%
\let\auto@bib@innerbib\@empty
\bibitem [{\citenamefont {Fukugita}\ and\ \citenamefont
  {Yanagida}(1986)}]{Fukugita:1986hr}%
  \BibitemOpen
  \bibfield  {author} {\bibinfo {author} {\bibfnamefont {M.}~\bibnamefont
  {Fukugita}}\ and\ \bibinfo {author} {\bibfnamefont {T.}~\bibnamefont
  {Yanagida}},\ }\bibfield  {title} {\enquote {\bibinfo {title} {{Baryogenesis
  Without Grand Unification}},}\ }\href {\doibase 10.1016/0370-2693(86)91126-3}
  {\bibfield  {journal} {\bibinfo  {journal} {Phys. Lett.}\ }\textbf {\bibinfo
  {volume} {B174}},\ \bibinfo {pages} {45--47} (\bibinfo {year}
  {1986})}\BibitemShut {NoStop}%
\bibitem [{\citenamefont {Fritzsch}\ and\ \citenamefont
  {Minkowski}(1975)}]{Fritzsch:1974nn}%
  \BibitemOpen
  \bibfield  {author} {\bibinfo {author} {\bibfnamefont {Harald}\ \bibnamefont
  {Fritzsch}}\ and\ \bibinfo {author} {\bibfnamefont {Peter}\ \bibnamefont
  {Minkowski}},\ }\bibfield  {title} {\enquote {\bibinfo {title} {{Unified
  Interactions of Leptons and Hadrons}},}\ }\href {\doibase
  10.1016/0003-4916(75)90211-0} {\bibfield  {journal} {\bibinfo  {journal}
  {Annals Phys.}\ }\textbf {\bibinfo {volume} {93}},\ \bibinfo {pages}
  {193--266} (\bibinfo {year} {1975})}\BibitemShut {NoStop}%
\bibitem [{\citenamefont {Gell-Mann}\ \emph {et~al.}(1979)\citenamefont
  {Gell-Mann}, \citenamefont {Ramond},\ and\ \citenamefont
  {Slansky}}]{GellMann:1980vs}%
  \BibitemOpen
  \bibfield  {author} {\bibinfo {author} {\bibfnamefont {Murray}\ \bibnamefont
  {Gell-Mann}}, \bibinfo {author} {\bibfnamefont {Pierre}\ \bibnamefont
  {Ramond}}, \ and\ \bibinfo {author} {\bibfnamefont {Richard}\ \bibnamefont
  {Slansky}},\ }\bibfield  {title} {\enquote {\bibinfo {title} {{Complex
  Spinors and Unified Theories}},}\ }\href@noop {} {\bibfield  {journal}
  {\bibinfo  {journal} {Conf.Proc.}\ }\textbf {\bibinfo {volume} {C790927}},\
  \bibinfo {pages} {315--321} (\bibinfo {year} {1979})},\ \Eprint
  {http://arxiv.org/abs/1306.4669} {arXiv:1306.4669 [hep-th]} \BibitemShut
  {NoStop}%
\bibitem [{\citenamefont {B\"odeker}\ and\ \citenamefont
  {Buchm\"uller}(2021)}]{Bodeker:2020ghk}%
  \BibitemOpen
  \bibfield  {author} {\bibinfo {author} {\bibfnamefont {Dietrich}\
  \bibnamefont {B\"odeker}}\ and\ \bibinfo {author} {\bibfnamefont {Wilfried}\
  \bibnamefont {Buchm\"uller}},\ }\bibfield  {title} {\enquote {\bibinfo
  {title} {{Baryogenesis from the weak scale to the grand unification
  scale}},}\ }\href {\doibase 10.1103/RevModPhys.93.035004} {\bibfield
  {journal} {\bibinfo  {journal} {Rev. Mod. Phys.}\ }\textbf {\bibinfo {volume}
  {93}},\ \bibinfo {pages} {035004} (\bibinfo {year} {2021})},\ \Eprint
  {http://arxiv.org/abs/2009.07294} {arXiv:2009.07294 [hep-ph]} \BibitemShut
  {NoStop}%
\bibitem [{\citenamefont {Di~Bari}(2022)}]{DiBari:2021fhs}%
  \BibitemOpen
  \bibfield  {author} {\bibinfo {author} {\bibfnamefont {Pasquale}\
  \bibnamefont {Di~Bari}},\ }\bibfield  {title} {\enquote {\bibinfo {title}
  {{On the origin of matter in the Universe}},}\ }\href {\doibase
  10.1016/j.ppnp.2021.103913} {\bibfield  {journal} {\bibinfo  {journal} {Prog.
  Part. Nucl. Phys.}\ }\textbf {\bibinfo {volume} {122}},\ \bibinfo {pages}
  {103913} (\bibinfo {year} {2022})},\ \Eprint
  {http://arxiv.org/abs/2107.13750} {arXiv:2107.13750 [hep-ph]} \BibitemShut
  {NoStop}%
\bibitem [{\citenamefont {Xing}\ and\ \citenamefont
  {Zhao}(2021)}]{Xing:2020ald}%
  \BibitemOpen
  \bibfield  {author} {\bibinfo {author} {\bibfnamefont {Zhi-zhong}\
  \bibnamefont {Xing}}\ and\ \bibinfo {author} {\bibfnamefont {Zhen-hua}\
  \bibnamefont {Zhao}},\ }\bibfield  {title} {\enquote {\bibinfo {title} {{The
  minimal seesaw and leptogenesis models}},}\ }\href {\doibase
  10.1088/1361-6633/abf086} {\bibfield  {journal} {\bibinfo  {journal} {Rept.
  Prog. Phys.}\ }\textbf {\bibinfo {volume} {84}},\ \bibinfo {pages} {066201}
  (\bibinfo {year} {2021})},\ \Eprint {http://arxiv.org/abs/2008.12090}
  {arXiv:2008.12090 [hep-ph]} \BibitemShut {NoStop}%
\bibitem [{\citenamefont {Buchmuller}\ and\ \citenamefont
  {Plumacher}(1996)}]{Buchmuller:1996pa}%
  \BibitemOpen
  \bibfield  {author} {\bibinfo {author} {\bibfnamefont {W.}~\bibnamefont
  {Buchmuller}}\ and\ \bibinfo {author} {\bibfnamefont {M.}~\bibnamefont
  {Plumacher}},\ }\bibfield  {title} {\enquote {\bibinfo {title} {{Baryon
  asymmetry and neutrino mixing}},}\ }\href {\doibase
  10.1016/S0370-2693(96)01232-4} {\bibfield  {journal} {\bibinfo  {journal}
  {Phys. Lett. B}\ }\textbf {\bibinfo {volume} {389}},\ \bibinfo {pages}
  {73--77} (\bibinfo {year} {1996})},\ \Eprint
  {http://arxiv.org/abs/hep-ph/9608308} {arXiv:hep-ph/9608308} \BibitemShut
  {NoStop}%
\bibitem [{\citenamefont {Nezri}\ and\ \citenamefont
  {Orloff}(2003)}]{Nezri:2000pb}%
  \BibitemOpen
  \bibfield  {author} {\bibinfo {author} {\bibfnamefont {Emmanuel}\
  \bibnamefont {Nezri}}\ and\ \bibinfo {author} {\bibfnamefont {Jean}\
  \bibnamefont {Orloff}},\ }\bibfield  {title} {\enquote {\bibinfo {title}
  {{Neutrino oscillations versus leptogenesis in SO(10) models}},}\ }\href
  {\doibase 10.1088/1126-6708/2003/04/020} {\bibfield  {journal} {\bibinfo
  {journal} {JHEP}\ }\textbf {\bibinfo {volume} {04}},\ \bibinfo {pages} {020}
  (\bibinfo {year} {2003})},\ \Eprint {http://arxiv.org/abs/hep-ph/0004227}
  {arXiv:hep-ph/0004227} \BibitemShut {NoStop}%
\bibitem [{\citenamefont {Buccella}\ \emph {et~al.}(2002)\citenamefont
  {Buccella}, \citenamefont {Falcone},\ and\ \citenamefont
  {Tramontano}}]{Buccella:2001tq}%
  \BibitemOpen
  \bibfield  {author} {\bibinfo {author} {\bibfnamefont {F.}~\bibnamefont
  {Buccella}}, \bibinfo {author} {\bibfnamefont {D.}~\bibnamefont {Falcone}}, \
  and\ \bibinfo {author} {\bibfnamefont {F.}~\bibnamefont {Tramontano}},\
  }\bibfield  {title} {\enquote {\bibinfo {title} {{Baryogenesis via
  leptogenesis in SO(10) models}},}\ }\href {\doibase
  10.1016/S0370-2693(01)01409-5} {\bibfield  {journal} {\bibinfo  {journal}
  {Phys. Lett. B}\ }\textbf {\bibinfo {volume} {524}},\ \bibinfo {pages}
  {241--244} (\bibinfo {year} {2002})},\ \Eprint
  {http://arxiv.org/abs/hep-ph/0108172} {arXiv:hep-ph/0108172} \BibitemShut
  {NoStop}%
\bibitem [{\citenamefont {Branco}\ \emph {et~al.}(2002)\citenamefont {Branco},
  \citenamefont {Gonzalez~Felipe}, \citenamefont {Joaquim},\ and\ \citenamefont
  {Rebelo}}]{Branco:2002kt}%
  \BibitemOpen
  \bibfield  {author} {\bibinfo {author} {\bibfnamefont {G.~C.}\ \bibnamefont
  {Branco}}, \bibinfo {author} {\bibfnamefont {R.}~\bibnamefont
  {Gonzalez~Felipe}}, \bibinfo {author} {\bibfnamefont {F.~R.}\ \bibnamefont
  {Joaquim}}, \ and\ \bibinfo {author} {\bibfnamefont {M.~N.}\ \bibnamefont
  {Rebelo}},\ }\bibfield  {title} {\enquote {\bibinfo {title} {{Leptogenesis,
  CP violation and neutrino data: What can we learn?}}}\ }\href {\doibase
  10.1016/S0550-3213(02)00478-9} {\bibfield  {journal} {\bibinfo  {journal}
  {Nucl. Phys. B}\ }\textbf {\bibinfo {volume} {640}},\ \bibinfo {pages}
  {202--232} (\bibinfo {year} {2002})},\ \Eprint
  {http://arxiv.org/abs/hep-ph/0202030} {arXiv:hep-ph/0202030} \BibitemShut
  {NoStop}%
\bibitem [{\citenamefont {Akhmedov}\ \emph {et~al.}(2003)\citenamefont
  {Akhmedov}, \citenamefont {Frigerio},\ and\ \citenamefont
  {Smirnov}}]{Akhmedov:2003dg}%
  \BibitemOpen
  \bibfield  {author} {\bibinfo {author} {\bibfnamefont {Evgeny~K.}\
  \bibnamefont {Akhmedov}}, \bibinfo {author} {\bibfnamefont {Michele}\
  \bibnamefont {Frigerio}}, \ and\ \bibinfo {author} {\bibfnamefont
  {Alexei~Yu.}\ \bibnamefont {Smirnov}},\ }\bibfield  {title} {\enquote
  {\bibinfo {title} {{Probing the seesaw mechanism with neutrino data and
  leptogenesis}},}\ }\href {\doibase 10.1088/1126-6708/2003/09/021} {\bibfield
  {journal} {\bibinfo  {journal} {JHEP}\ }\textbf {\bibinfo {volume} {09}},\
  \bibinfo {pages} {021} (\bibinfo {year} {2003})},\ \Eprint
  {http://arxiv.org/abs/hep-ph/0305322} {arXiv:hep-ph/0305322} \BibitemShut
  {NoStop}%
\bibitem [{\citenamefont {Di~Bari}\ and\ \citenamefont
  {Riotto}(2009)}]{DiBari:2008mp}%
  \BibitemOpen
  \bibfield  {author} {\bibinfo {author} {\bibfnamefont {Pasquale}\
  \bibnamefont {Di~Bari}}\ and\ \bibinfo {author} {\bibfnamefont {Antonio}\
  \bibnamefont {Riotto}},\ }\bibfield  {title} {\enquote {\bibinfo {title}
  {{Successful type I Leptogenesis with SO(10)-inspired mass relations}},}\
  }\href {\doibase 10.1016/j.physletb.2008.12.054} {\bibfield  {journal}
  {\bibinfo  {journal} {Phys. Lett. B}\ }\textbf {\bibinfo {volume} {671}},\
  \bibinfo {pages} {462--469} (\bibinfo {year} {2009})},\ \Eprint
  {http://arxiv.org/abs/0809.2285} {arXiv:0809.2285 [hep-ph]} \BibitemShut
  {NoStop}%
\bibitem [{\citenamefont {Di~Bari}\ and\ \citenamefont
  {Riotto}(2011)}]{DiBari:2010ux}%
  \BibitemOpen
  \bibfield  {author} {\bibinfo {author} {\bibfnamefont {Pasquale}\
  \bibnamefont {Di~Bari}}\ and\ \bibinfo {author} {\bibfnamefont {Antonio}\
  \bibnamefont {Riotto}},\ }\bibfield  {title} {\enquote {\bibinfo {title}
  {{Testing SO(10)-inspired leptogenesis with low energy neutrino
  experiments}},}\ }\href {\doibase 10.1088/1475-7516/2011/04/037} {\bibfield
  {journal} {\bibinfo  {journal} {JCAP}\ }\textbf {\bibinfo {volume} {04}},\
  \bibinfo {pages} {037} (\bibinfo {year} {2011})},\ \Eprint
  {http://arxiv.org/abs/1012.2343} {arXiv:1012.2343 [hep-ph]} \BibitemShut
  {NoStop}%
\bibitem [{\citenamefont {Buccella}\ \emph {et~al.}(2012)\citenamefont
  {Buccella}, \citenamefont {Falcone}, \citenamefont {Fong}, \citenamefont
  {Nardi},\ and\ \citenamefont {Ricciardi}}]{Buccella:2012kc}%
  \BibitemOpen
  \bibfield  {author} {\bibinfo {author} {\bibfnamefont {Franco}\ \bibnamefont
  {Buccella}}, \bibinfo {author} {\bibfnamefont {Domenico}\ \bibnamefont
  {Falcone}}, \bibinfo {author} {\bibfnamefont {Chee~Sheng}\ \bibnamefont
  {Fong}}, \bibinfo {author} {\bibfnamefont {Enrico}\ \bibnamefont {Nardi}}, \
  and\ \bibinfo {author} {\bibfnamefont {Giulia}\ \bibnamefont {Ricciardi}},\
  }\bibfield  {title} {\enquote {\bibinfo {title} {{Squeezing out predictions
  with leptogenesis from SO(10)}},}\ }\href {\doibase
  10.1103/PhysRevD.86.035012} {\bibfield  {journal} {\bibinfo  {journal} {Phys.
  Rev. D}\ }\textbf {\bibinfo {volume} {86}},\ \bibinfo {pages} {035012}
  (\bibinfo {year} {2012})},\ \Eprint {http://arxiv.org/abs/1203.0829}
  {arXiv:1203.0829 [hep-ph]} \BibitemShut {NoStop}%
\bibitem [{\citenamefont {Di~Bari}\ \emph {et~al.}(2015)\citenamefont
  {Di~Bari}, \citenamefont {Marzola},\ and\ \citenamefont
  {Re~Fiorentin}}]{DiBari:2014eya}%
  \BibitemOpen
  \bibfield  {author} {\bibinfo {author} {\bibfnamefont {Pasquale}\
  \bibnamefont {Di~Bari}}, \bibinfo {author} {\bibfnamefont {Luca}\
  \bibnamefont {Marzola}}, \ and\ \bibinfo {author} {\bibfnamefont {Michele}\
  \bibnamefont {Re~Fiorentin}},\ }\bibfield  {title} {\enquote {\bibinfo
  {title} {{Decrypting $SO(10)$-inspired leptogenesis}},}\ }\href {\doibase
  10.1016/j.nuclphysb.2015.02.005} {\bibfield  {journal} {\bibinfo  {journal}
  {Nucl. Phys. B}\ }\textbf {\bibinfo {volume} {893}},\ \bibinfo {pages}
  {122--157} (\bibinfo {year} {2015})},\ \Eprint
  {http://arxiv.org/abs/1411.5478} {arXiv:1411.5478 [hep-ph]} \BibitemShut
  {NoStop}%
\bibitem [{\citenamefont {Di~Bari}\ and\ \citenamefont
  {King}(2015)}]{DiBari:2015oca}%
  \BibitemOpen
  \bibfield  {author} {\bibinfo {author} {\bibfnamefont {Pasquale}\
  \bibnamefont {Di~Bari}}\ and\ \bibinfo {author} {\bibfnamefont {Stephen~F.}\
  \bibnamefont {King}},\ }\bibfield  {title} {\enquote {\bibinfo {title}
  {{Successful $N_2$ leptogenesis with flavour coupling effects in realistic
  unified models}},}\ }\href {\doibase 10.1088/1475-7516/2015/10/008}
  {\bibfield  {journal} {\bibinfo  {journal} {JCAP}\ }\textbf {\bibinfo
  {volume} {10}},\ \bibinfo {pages} {008} (\bibinfo {year} {2015})},\ \Eprint
  {http://arxiv.org/abs/1507.06431} {arXiv:1507.06431 [hep-ph]} \BibitemShut
  {NoStop}%
\bibitem [{\citenamefont {Di~Bari}\ and\ \citenamefont
  {Re~Fiorentin}(2017)}]{DiBari:2017uka}%
  \BibitemOpen
  \bibfield  {author} {\bibinfo {author} {\bibfnamefont {Pasquale}\
  \bibnamefont {Di~Bari}}\ and\ \bibinfo {author} {\bibfnamefont {Michele}\
  \bibnamefont {Re~Fiorentin}},\ }\bibfield  {title} {\enquote {\bibinfo
  {title} {{A full analytic solution of $SO(10)$-inspired leptogenesis}},}\
  }\href {\doibase 10.1007/JHEP10(2017)029} {\bibfield  {journal} {\bibinfo
  {journal} {JHEP}\ }\textbf {\bibinfo {volume} {10}},\ \bibinfo {pages} {029}
  (\bibinfo {year} {2017})},\ \Eprint {http://arxiv.org/abs/1705.01935}
  {arXiv:1705.01935 [hep-ph]} \BibitemShut {NoStop}%
\bibitem [{\citenamefont {Di~Bari}\ and\ \citenamefont
  {Samanta}(2020)}]{DiBari:2020plh}%
  \BibitemOpen
  \bibfield  {author} {\bibinfo {author} {\bibfnamefont {Pasquale}\
  \bibnamefont {Di~Bari}}\ and\ \bibinfo {author} {\bibfnamefont {Rome}\
  \bibnamefont {Samanta}},\ }\bibfield  {title} {\enquote {\bibinfo {title}
  {{The $SO(10)$-inspired leptogenesis timely opportunity}},}\ }\href {\doibase
  10.1007/JHEP08(2020)124} {\bibfield  {journal} {\bibinfo  {journal} {JHEP}\
  }\textbf {\bibinfo {volume} {08}},\ \bibinfo {pages} {124} (\bibinfo {year}
  {2020})},\ \Eprint {http://arxiv.org/abs/2005.03057} {arXiv:2005.03057
  [hep-ph]} \BibitemShut {NoStop}%
\bibitem [{\citenamefont {Fong}\ \emph {et~al.}(2015)\citenamefont {Fong},
  \citenamefont {Meloni}, \citenamefont {Meroni},\ and\ \citenamefont
  {Nardi}}]{Fong:2014gea}%
  \BibitemOpen
  \bibfield  {author} {\bibinfo {author} {\bibfnamefont {Chee~Sheng}\
  \bibnamefont {Fong}}, \bibinfo {author} {\bibfnamefont {Davide}\ \bibnamefont
  {Meloni}}, \bibinfo {author} {\bibfnamefont {Aurora}\ \bibnamefont {Meroni}},
  \ and\ \bibinfo {author} {\bibfnamefont {Enrico}\ \bibnamefont {Nardi}},\
  }\bibfield  {title} {\enquote {\bibinfo {title} {{Leptogenesis in SO(10)}},}\
  }\href {\doibase 10.1007/JHEP01(2015)111} {\bibfield  {journal} {\bibinfo
  {journal} {JHEP}\ }\textbf {\bibinfo {volume} {01}},\ \bibinfo {pages} {111}
  (\bibinfo {year} {2015})},\ \Eprint {http://arxiv.org/abs/1412.4776}
  {arXiv:1412.4776 [hep-ph]} \BibitemShut {NoStop}%
\bibitem [{\citenamefont {Mummidi}\ and\ \citenamefont
  {Patel}(2021)}]{Mummidi:2021anm}%
  \BibitemOpen
  \bibfield  {author} {\bibinfo {author} {\bibfnamefont {V.~Suryanarayana}\
  \bibnamefont {Mummidi}}\ and\ \bibinfo {author} {\bibfnamefont {Ketan~M.}\
  \bibnamefont {Patel}},\ }\bibfield  {title} {\enquote {\bibinfo {title}
  {{Leptogenesis and fermion mass fit in a renormalizable SO(10) model}},}\
  }\href {\doibase 10.1007/JHEP12(2021)042} {\bibfield  {journal} {\bibinfo
  {journal} {JHEP}\ }\textbf {\bibinfo {volume} {12}},\ \bibinfo {pages} {042}
  (\bibinfo {year} {2021})},\ \Eprint {http://arxiv.org/abs/2109.04050}
  {arXiv:2109.04050 [hep-ph]} \BibitemShut {NoStop}%
\bibitem [{\citenamefont {Patel}(2023)}]{Patel:2022xxu}%
  \BibitemOpen
  \bibfield  {author} {\bibinfo {author} {\bibfnamefont {Ketan~M.}\
  \bibnamefont {Patel}},\ }\bibfield  {title} {\enquote {\bibinfo {title}
  {{Minimal spontaneous CP-violating GUT and predictions for leptonic CP
  phases}},}\ }\href {\doibase 10.1103/PhysRevD.107.075041} {\bibfield
  {journal} {\bibinfo  {journal} {Phys. Rev. D}\ }\textbf {\bibinfo {volume}
  {107}},\ \bibinfo {pages} {075041} (\bibinfo {year} {2023})},\ \Eprint
  {http://arxiv.org/abs/2212.04095} {arXiv:2212.04095 [hep-ph]} \BibitemShut
  {NoStop}%
\bibitem [{\citenamefont {Kaladharan}\ and\ \citenamefont
  {Saad}(2024)}]{Kaladharan:2023zbr}%
  \BibitemOpen
  \bibfield  {author} {\bibinfo {author} {\bibfnamefont {Ajay}\ \bibnamefont
  {Kaladharan}}\ and\ \bibinfo {author} {\bibfnamefont {Shaikh}\ \bibnamefont
  {Saad}},\ }\bibfield  {title} {\enquote {\bibinfo {title} {{Fermion mass,
  axion dark matter, and leptogenesis in SO(10) GUT}},}\ }\href {\doibase
  10.1103/PhysRevD.109.055010} {\bibfield  {journal} {\bibinfo  {journal}
  {Phys. Rev. D}\ }\textbf {\bibinfo {volume} {109}},\ \bibinfo {pages}
  {055010} (\bibinfo {year} {2024})},\ \Eprint
  {http://arxiv.org/abs/2308.04497} {arXiv:2308.04497 [hep-ph]} \BibitemShut
  {NoStop}%
\bibitem [{\citenamefont {Babu}\ \emph {et~al.}(2024)\citenamefont {Babu},
  \citenamefont {Di~Bari}, \citenamefont {Fong},\ and\ \citenamefont
  {Saad}}]{Babu:2024ahk}%
  \BibitemOpen
  \bibfield  {author} {\bibinfo {author} {\bibfnamefont {K.~S.}\ \bibnamefont
  {Babu}}, \bibinfo {author} {\bibfnamefont {Pasquale}\ \bibnamefont
  {Di~Bari}}, \bibinfo {author} {\bibfnamefont {Chee~Sheng}\ \bibnamefont
  {Fong}}, \ and\ \bibinfo {author} {\bibfnamefont {Shaikh}\ \bibnamefont
  {Saad}},\ }\bibfield  {title} {\enquote {\bibinfo {title} {{Leptogenesis in
  SO(10) with minimal Yukawa sector}},}\ }\href {\doibase
  10.1007/JHEP10(2024)190} {\bibfield  {journal} {\bibinfo  {journal} {JHEP}\
  }\textbf {\bibinfo {volume} {10}},\ \bibinfo {pages} {190} (\bibinfo {year}
  {2024})},\ \Eprint {http://arxiv.org/abs/2409.03840} {arXiv:2409.03840
  [hep-ph]} \BibitemShut {NoStop}%
\bibitem [{\citenamefont {Babu}\ and\ \citenamefont
  {Mohapatra}(1993)}]{Babu:1992ia}%
  \BibitemOpen
  \bibfield  {author} {\bibinfo {author} {\bibfnamefont {K.~S.}\ \bibnamefont
  {Babu}}\ and\ \bibinfo {author} {\bibfnamefont {R.~N.}\ \bibnamefont
  {Mohapatra}},\ }\bibfield  {title} {\enquote {\bibinfo {title} {{Predictive
  neutrino spectrum in minimal SO(10) grand unification}},}\ }\href {\doibase
  10.1103/PhysRevLett.70.2845} {\bibfield  {journal} {\bibinfo  {journal}
  {Phys. Rev. Lett.}\ }\textbf {\bibinfo {volume} {70}},\ \bibinfo {pages}
  {2845--2848} (\bibinfo {year} {1993})},\ \Eprint
  {http://arxiv.org/abs/hep-ph/9209215} {arXiv:hep-ph/9209215 [hep-ph]}
  \BibitemShut {NoStop}%
\bibitem [{\citenamefont {Bajc}\ \emph {et~al.}(2006)\citenamefont {Bajc},
  \citenamefont {Melfo}, \citenamefont {Senjanovic},\ and\ \citenamefont
  {Vissani}}]{Bajc:2005zf}%
  \BibitemOpen
  \bibfield  {author} {\bibinfo {author} {\bibfnamefont {Borut}\ \bibnamefont
  {Bajc}}, \bibinfo {author} {\bibfnamefont {Alejandra}\ \bibnamefont {Melfo}},
  \bibinfo {author} {\bibfnamefont {Goran}\ \bibnamefont {Senjanovic}}, \ and\
  \bibinfo {author} {\bibfnamefont {Francesco}\ \bibnamefont {Vissani}},\
  }\bibfield  {title} {\enquote {\bibinfo {title} {{Yukawa sector in
  non-supersymmetric renormalizable SO(10)}},}\ }\href {\doibase
  10.1103/PhysRevD.73.055001} {\bibfield  {journal} {\bibinfo  {journal} {Phys.
  Rev. D}\ }\textbf {\bibinfo {volume} {73}},\ \bibinfo {pages} {055001}
  (\bibinfo {year} {2006})},\ \Eprint {http://arxiv.org/abs/hep-ph/0510139}
  {arXiv:hep-ph/0510139} \BibitemShut {NoStop}%
\bibitem [{\citenamefont {Joshipura}\ and\ \citenamefont
  {Patel}(2011)}]{Joshipura:2011nn}%
  \BibitemOpen
  \bibfield  {author} {\bibinfo {author} {\bibfnamefont {Anjan~S.}\
  \bibnamefont {Joshipura}}\ and\ \bibinfo {author} {\bibfnamefont {Ketan~M.}\
  \bibnamefont {Patel}},\ }\bibfield  {title} {\enquote {\bibinfo {title}
  {{Fermion Masses in SO(10) Models}},}\ }\href {\doibase
  10.1103/PhysRevD.83.095002} {\bibfield  {journal} {\bibinfo  {journal} {Phys.
  Rev. D}\ }\textbf {\bibinfo {volume} {83}},\ \bibinfo {pages} {095002}
  (\bibinfo {year} {2011})},\ \Eprint {http://arxiv.org/abs/1102.5148}
  {arXiv:1102.5148 [hep-ph]} \BibitemShut {NoStop}%
\bibitem [{\citenamefont {Altarelli}\ and\ \citenamefont
  {Meloni}(2013)}]{Altarelli:2013aqa}%
  \BibitemOpen
  \bibfield  {author} {\bibinfo {author} {\bibfnamefont {Guido}\ \bibnamefont
  {Altarelli}}\ and\ \bibinfo {author} {\bibfnamefont {Davide}\ \bibnamefont
  {Meloni}},\ }\bibfield  {title} {\enquote {\bibinfo {title} {{A non
  supersymmetric SO(10) grand unified model for all the physics below
  $M_{GUT}$}},}\ }\href {\doibase 10.1007/JHEP08(2013)021} {\bibfield
  {journal} {\bibinfo  {journal} {JHEP}\ }\textbf {\bibinfo {volume} {08}},\
  \bibinfo {pages} {021} (\bibinfo {year} {2013})},\ \Eprint
  {http://arxiv.org/abs/1305.1001} {arXiv:1305.1001 [hep-ph]} \BibitemShut
  {NoStop}%
\bibitem [{\citenamefont {Dueck}\ and\ \citenamefont
  {Rodejohann}(2013)}]{Dueck:2013gca}%
  \BibitemOpen
  \bibfield  {author} {\bibinfo {author} {\bibfnamefont {Alexander}\
  \bibnamefont {Dueck}}\ and\ \bibinfo {author} {\bibfnamefont {Werner}\
  \bibnamefont {Rodejohann}},\ }\bibfield  {title} {\enquote {\bibinfo {title}
  {{Fits to SO(10) Grand Unified Models}},}\ }\href {\doibase
  10.1007/JHEP09(2013)024} {\bibfield  {journal} {\bibinfo  {journal} {JHEP}\
  }\textbf {\bibinfo {volume} {09}},\ \bibinfo {pages} {024} (\bibinfo {year}
  {2013})},\ \Eprint {http://arxiv.org/abs/1306.4468} {arXiv:1306.4468
  [hep-ph]} \BibitemShut {NoStop}%
\bibitem [{\citenamefont {Meloni}\ \emph {et~al.}(2017)\citenamefont {Meloni},
  \citenamefont {Ohlsson},\ and\ \citenamefont {Riad}}]{Meloni:2016rnt}%
  \BibitemOpen
  \bibfield  {author} {\bibinfo {author} {\bibfnamefont {Davide}\ \bibnamefont
  {Meloni}}, \bibinfo {author} {\bibfnamefont {Tommy}\ \bibnamefont {Ohlsson}},
  \ and\ \bibinfo {author} {\bibfnamefont {Stella}\ \bibnamefont {Riad}},\
  }\bibfield  {title} {\enquote {\bibinfo {title} {{Renormalization Group
  Running of Fermion Observables in an Extended Non-Supersymmetric SO(10)
  Model}},}\ }\href {\doibase 10.1007/JHEP03(2017)045} {\bibfield  {journal}
  {\bibinfo  {journal} {JHEP}\ }\textbf {\bibinfo {volume} {03}},\ \bibinfo
  {pages} {045} (\bibinfo {year} {2017})},\ \Eprint
  {http://arxiv.org/abs/1612.07973} {arXiv:1612.07973 [hep-ph]} \BibitemShut
  {NoStop}%
\bibitem [{\citenamefont {Babu}\ \emph {et~al.}(2017)\citenamefont {Babu},
  \citenamefont {Bajc},\ and\ \citenamefont {Saad}}]{Babu:2016bmy}%
  \BibitemOpen
  \bibfield  {author} {\bibinfo {author} {\bibfnamefont {K.~S.}\ \bibnamefont
  {Babu}}, \bibinfo {author} {\bibfnamefont {Borut}\ \bibnamefont {Bajc}}, \
  and\ \bibinfo {author} {\bibfnamefont {Shaikh}\ \bibnamefont {Saad}},\
  }\bibfield  {title} {\enquote {\bibinfo {title} {{Yukawa Sector of Minimal
  SO(10) Unification}},}\ }\href {\doibase 10.1007/JHEP02(2017)136} {\bibfield
  {journal} {\bibinfo  {journal} {JHEP}\ }\textbf {\bibinfo {volume} {02}},\
  \bibinfo {pages} {136} (\bibinfo {year} {2017})},\ \Eprint
  {http://arxiv.org/abs/1612.04329} {arXiv:1612.04329 [hep-ph]} \BibitemShut
  {NoStop}%
\bibitem [{\citenamefont {Ohlsson}\ and\ \citenamefont
  {Pernow}(2019)}]{Ohlsson:2019sja}%
  \BibitemOpen
  \bibfield  {author} {\bibinfo {author} {\bibfnamefont {Tommy}\ \bibnamefont
  {Ohlsson}}\ and\ \bibinfo {author} {\bibfnamefont {Marcus}\ \bibnamefont
  {Pernow}},\ }\bibfield  {title} {\enquote {\bibinfo {title} {{Fits to
  Non-Supersymmetric SO(10) Models with Type I and II Seesaw Mechanisms Using
  Renormalization Group Evolution}},}\ }\href {\doibase
  10.1007/JHEP06(2019)085} {\bibfield  {journal} {\bibinfo  {journal} {JHEP}\
  }\textbf {\bibinfo {volume} {06}},\ \bibinfo {pages} {085} (\bibinfo {year}
  {2019})},\ \Eprint {http://arxiv.org/abs/1903.08241} {arXiv:1903.08241
  [hep-ph]} \BibitemShut {NoStop}%
\bibitem [{\citenamefont {Chun}\ and\ \citenamefont
  {Kang}(2001)}]{Chun:2000dr}%
  \BibitemOpen
  \bibfield  {author} {\bibinfo {author} {\bibfnamefont {Eung~Jin}\
  \bibnamefont {Chun}}\ and\ \bibinfo {author} {\bibfnamefont {Sin~Kyu}\
  \bibnamefont {Kang}},\ }\bibfield  {title} {\enquote {\bibinfo {title}
  {{Baryogenesis and degenerate neutrinos}},}\ }\href {\doibase
  10.1103/PhysRevD.63.097902} {\bibfield  {journal} {\bibinfo  {journal} {Phys.
  Rev. D}\ }\textbf {\bibinfo {volume} {63}},\ \bibinfo {pages} {097902}
  (\bibinfo {year} {2001})},\ \Eprint {http://arxiv.org/abs/hep-ph/0001296}
  {arXiv:hep-ph/0001296} \BibitemShut {NoStop}%
\bibitem [{\citenamefont {Joshipura}\ \emph {et~al.}(2001)\citenamefont
  {Joshipura}, \citenamefont {Paschos},\ and\ \citenamefont
  {Rodejohann}}]{Joshipura:2001ya}%
  \BibitemOpen
  \bibfield  {author} {\bibinfo {author} {\bibfnamefont {Anjan~S.}\
  \bibnamefont {Joshipura}}, \bibinfo {author} {\bibfnamefont {Emmanuel~A.}\
  \bibnamefont {Paschos}}, \ and\ \bibinfo {author} {\bibfnamefont {Werner}\
  \bibnamefont {Rodejohann}},\ }\bibfield  {title} {\enquote {\bibinfo {title}
  {{Leptogenesis in left-right symmetric theories}},}\ }\href {\doibase
  10.1016/S0550-3213(01)00346-7} {\bibfield  {journal} {\bibinfo  {journal}
  {Nucl. Phys. B}\ }\textbf {\bibinfo {volume} {611}},\ \bibinfo {pages}
  {227--238} (\bibinfo {year} {2001})},\ \Eprint
  {http://arxiv.org/abs/hep-ph/0104228} {arXiv:hep-ph/0104228} \BibitemShut
  {NoStop}%
\bibitem [{\citenamefont {Antusch}\ and\ \citenamefont
  {King}(2004)}]{Antusch:2004xy}%
  \BibitemOpen
  \bibfield  {author} {\bibinfo {author} {\bibfnamefont {Stefan}\ \bibnamefont
  {Antusch}}\ and\ \bibinfo {author} {\bibfnamefont {Steve~F.}\ \bibnamefont
  {King}},\ }\bibfield  {title} {\enquote {\bibinfo {title} {{Type II
  Leptogenesis and the neutrino mass scale}},}\ }\href {\doibase
  10.1016/j.physletb.2004.07.009} {\bibfield  {journal} {\bibinfo  {journal}
  {Phys. Lett. B}\ }\textbf {\bibinfo {volume} {597}},\ \bibinfo {pages}
  {199--207} (\bibinfo {year} {2004})},\ \Eprint
  {http://arxiv.org/abs/hep-ph/0405093} {arXiv:hep-ph/0405093} \BibitemShut
  {NoStop}%
\bibitem [{\citenamefont {Antusch}\ and\ \citenamefont
  {King}(2006)}]{Antusch:2005tu}%
  \BibitemOpen
  \bibfield  {author} {\bibinfo {author} {\bibfnamefont {Stefan}\ \bibnamefont
  {Antusch}}\ and\ \bibinfo {author} {\bibfnamefont {Steve~F.}\ \bibnamefont
  {King}},\ }\bibfield  {title} {\enquote {\bibinfo {title} {{Leptogenesis in
  unified theories with type II see-saw}},}\ }\href {\doibase
  10.1088/1126-6708/2006/01/117} {\bibfield  {journal} {\bibinfo  {journal}
  {JHEP}\ }\textbf {\bibinfo {volume} {01}},\ \bibinfo {pages} {117} (\bibinfo
  {year} {2006})},\ \Eprint {http://arxiv.org/abs/hep-ph/0507333}
  {arXiv:hep-ph/0507333} \BibitemShut {NoStop}%
\bibitem [{\citenamefont {Aristizabal~Sierra}\ \emph
  {et~al.}(2012)\citenamefont {Aristizabal~Sierra}, \citenamefont {Bazzocchi},\
  and\ \citenamefont {de~Medeiros~Varzielas}}]{AristizabalSierra:2011ab}%
  \BibitemOpen
  \bibfield  {author} {\bibinfo {author} {\bibfnamefont {D.}~\bibnamefont
  {Aristizabal~Sierra}}, \bibinfo {author} {\bibfnamefont {F.}~\bibnamefont
  {Bazzocchi}}, \ and\ \bibinfo {author} {\bibfnamefont {I.}~\bibnamefont
  {de~Medeiros~Varzielas}},\ }\bibfield  {title} {\enquote {\bibinfo {title}
  {{Leptogenesis in flavor models with type I and II seesaws}},}\ }\href
  {\doibase 10.1016/j.nuclphysb.2012.01.009} {\bibfield  {journal} {\bibinfo
  {journal} {Nucl. Phys. B}\ }\textbf {\bibinfo {volume} {858}},\ \bibinfo
  {pages} {196--213} (\bibinfo {year} {2012})},\ \Eprint
  {http://arxiv.org/abs/1112.1843} {arXiv:1112.1843 [hep-ph]} \BibitemShut
  {NoStop}%
\bibitem [{\citenamefont {Pramanick}\ \emph {et~al.}(2023)\citenamefont
  {Pramanick}, \citenamefont {Ray},\ and\ \citenamefont
  {Shaw}}]{Pramanick:2022put}%
  \BibitemOpen
  \bibfield  {author} {\bibinfo {author} {\bibfnamefont {Rohan}\ \bibnamefont
  {Pramanick}}, \bibinfo {author} {\bibfnamefont {Tirtha~Sankar}\ \bibnamefont
  {Ray}}, \ and\ \bibinfo {author} {\bibfnamefont {Avirup}\ \bibnamefont
  {Shaw}},\ }\bibfield  {title} {\enquote {\bibinfo {title} {{Neutrino mass and
  leptogenesis in a hybrid seesaw model with a spontaneously broken CP}},}\
  }\href {\doibase 10.1007/JHEP06(2023)099} {\bibfield  {journal} {\bibinfo
  {journal} {JHEP}\ }\textbf {\bibinfo {volume} {06}},\ \bibinfo {pages} {099}
  (\bibinfo {year} {2023})},\ \Eprint {http://arxiv.org/abs/2211.04403}
  {arXiv:2211.04403 [hep-ph]} \BibitemShut {NoStop}%
\bibitem [{\citenamefont {Pramanick}\ \emph {et~al.}(2024)\citenamefont
  {Pramanick}, \citenamefont {Ray},\ and\ \citenamefont
  {Sil}}]{Pramanick:2024gvu}%
  \BibitemOpen
  \bibfield  {author} {\bibinfo {author} {\bibfnamefont {Rohan}\ \bibnamefont
  {Pramanick}}, \bibinfo {author} {\bibfnamefont {Tirtha~Sankar}\ \bibnamefont
  {Ray}}, \ and\ \bibinfo {author} {\bibfnamefont {Arunansu}\ \bibnamefont
  {Sil}},\ }\bibfield  {title} {\enquote {\bibinfo {title} {{Toward a more
  complete description of hybrid leptogenesis}},}\ }\href {\doibase
  10.1103/PhysRevD.109.115011} {\bibfield  {journal} {\bibinfo  {journal}
  {Phys. Rev. D}\ }\textbf {\bibinfo {volume} {109}},\ \bibinfo {pages}
  {115011} (\bibinfo {year} {2024})},\ \Eprint
  {http://arxiv.org/abs/2401.12189} {arXiv:2401.12189 [hep-ph]} \BibitemShut
  {NoStop}%
\bibitem [{\citenamefont {Peccei}\ and\ \citenamefont
  {Quinn}(1977)}]{Peccei:1977hh}%
  \BibitemOpen
  \bibfield  {author} {\bibinfo {author} {\bibfnamefont {R.~D.}\ \bibnamefont
  {Peccei}}\ and\ \bibinfo {author} {\bibfnamefont {Helen~R.}\ \bibnamefont
  {Quinn}},\ }\bibfield  {title} {\enquote {\bibinfo {title} {{CP Conservation
  in the Presence of Instantons}},}\ }\href {\doibase
  10.1103/PhysRevLett.38.1440} {\bibfield  {journal} {\bibinfo  {journal}
  {Phys. Rev. Lett.}\ }\textbf {\bibinfo {volume} {38}},\ \bibinfo {pages}
  {1440--1443} (\bibinfo {year} {1977})}\BibitemShut {NoStop}%
\bibitem [{\citenamefont {Wilczek}\ and\ \citenamefont
  {Zee}(1982)}]{Wilczek:1981iz}%
  \BibitemOpen
  \bibfield  {author} {\bibinfo {author} {\bibfnamefont {Frank}\ \bibnamefont
  {Wilczek}}\ and\ \bibinfo {author} {\bibfnamefont {A.}~\bibnamefont {Zee}},\
  }\bibfield  {title} {\enquote {\bibinfo {title} {{Families from Spinors}},}\
  }\href {\doibase 10.1103/PhysRevD.25.553} {\bibfield  {journal} {\bibinfo
  {journal} {Phys. Rev. D}\ }\textbf {\bibinfo {volume} {25}},\ \bibinfo
  {pages} {553} (\bibinfo {year} {1982})}\BibitemShut {NoStop}%
\bibitem [{\citenamefont {Bertolini}\ \emph {et~al.}(2009)\citenamefont
  {Bertolini}, \citenamefont {Di~Luzio},\ and\ \citenamefont
  {Malinsky}}]{Bertolini:2009qj}%
  \BibitemOpen
  \bibfield  {author} {\bibinfo {author} {\bibfnamefont {Stefano}\ \bibnamefont
  {Bertolini}}, \bibinfo {author} {\bibfnamefont {Luca}\ \bibnamefont
  {Di~Luzio}}, \ and\ \bibinfo {author} {\bibfnamefont {Michal}\ \bibnamefont
  {Malinsky}},\ }\bibfield  {title} {\enquote {\bibinfo {title} {{Intermediate
  mass scales in the non-supersymmetric SO(10) grand unification: A
  Reappraisal}},}\ }\href {\doibase 10.1103/PhysRevD.80.015013} {\bibfield
  {journal} {\bibinfo  {journal} {Phys. Rev. D}\ }\textbf {\bibinfo {volume}
  {80}},\ \bibinfo {pages} {015013} (\bibinfo {year} {2009})},\ \Eprint
  {http://arxiv.org/abs/0903.4049} {arXiv:0903.4049 [hep-ph]} \BibitemShut
  {NoStop}%
\bibitem [{\citenamefont {Covi}\ \emph {et~al.}(1996)\citenamefont {Covi},
  \citenamefont {Roulet},\ and\ \citenamefont {Vissani}}]{Covi:1996wh}%
  \BibitemOpen
  \bibfield  {author} {\bibinfo {author} {\bibfnamefont {Laura}\ \bibnamefont
  {Covi}}, \bibinfo {author} {\bibfnamefont {Esteban}\ \bibnamefont {Roulet}},
  \ and\ \bibinfo {author} {\bibfnamefont {Francesco}\ \bibnamefont
  {Vissani}},\ }\bibfield  {title} {\enquote {\bibinfo {title} {{CP violating
  decays in leptogenesis scenarios}},}\ }\href {\doibase
  10.1016/0370-2693(96)00817-9} {\bibfield  {journal} {\bibinfo  {journal}
  {Phys. Lett. B}\ }\textbf {\bibinfo {volume} {384}},\ \bibinfo {pages}
  {169--174} (\bibinfo {year} {1996})},\ \Eprint
  {http://arxiv.org/abs/hep-ph/9605319} {arXiv:hep-ph/9605319} \BibitemShut
  {NoStop}%
\bibitem [{\citenamefont {Hambye}\ and\ \citenamefont
  {Senjanovic}(2004)}]{Hambye:2003ka}%
  \BibitemOpen
  \bibfield  {author} {\bibinfo {author} {\bibfnamefont {Thomas}\ \bibnamefont
  {Hambye}}\ and\ \bibinfo {author} {\bibfnamefont {Goran}\ \bibnamefont
  {Senjanovic}},\ }\bibfield  {title} {\enquote {\bibinfo {title}
  {{Consequences of triplet seesaw for leptogenesis}},}\ }\href {\doibase
  10.1016/j.physletb.2003.11.061} {\bibfield  {journal} {\bibinfo  {journal}
  {Phys. Lett. B}\ }\textbf {\bibinfo {volume} {582}},\ \bibinfo {pages}
  {73--81} (\bibinfo {year} {2004})},\ \Eprint
  {http://arxiv.org/abs/hep-ph/0307237} {arXiv:hep-ph/0307237} \BibitemShut
  {NoStop}%
\bibitem [{\citenamefont {Blanchet}\ \emph {et~al.}(2013)\citenamefont
  {Blanchet}, \citenamefont {Di~Bari}, \citenamefont {Jones},\ and\
  \citenamefont {Marzola}}]{Blanchet:2011xq}%
  \BibitemOpen
  \bibfield  {author} {\bibinfo {author} {\bibfnamefont {Steve}\ \bibnamefont
  {Blanchet}}, \bibinfo {author} {\bibfnamefont {Pasquale}\ \bibnamefont
  {Di~Bari}}, \bibinfo {author} {\bibfnamefont {David~A.}\ \bibnamefont
  {Jones}}, \ and\ \bibinfo {author} {\bibfnamefont {Luca}\ \bibnamefont
  {Marzola}},\ }\bibfield  {title} {\enquote {\bibinfo {title} {{Leptogenesis
  with heavy neutrino flavours: from density matrix to Boltzmann equations}},}\
  }\href {\doibase 10.1088/1475-7516/2013/01/041} {\bibfield  {journal}
  {\bibinfo  {journal} {JCAP}\ }\textbf {\bibinfo {volume} {01}},\ \bibinfo
  {pages} {041} (\bibinfo {year} {2013})},\ \Eprint
  {http://arxiv.org/abs/1112.4528} {arXiv:1112.4528 [hep-ph]} \BibitemShut
  {NoStop}%
\bibitem [{\citenamefont {Aristizabal~Sierra}\ \emph
  {et~al.}(2014)\citenamefont {Aristizabal~Sierra}, \citenamefont {Dhen},\ and\
  \citenamefont {Hambye}}]{AristizabalSierra:2014nzr}%
  \BibitemOpen
  \bibfield  {author} {\bibinfo {author} {\bibfnamefont {D.}~\bibnamefont
  {Aristizabal~Sierra}}, \bibinfo {author} {\bibfnamefont {Mika\"el}\
  \bibnamefont {Dhen}}, \ and\ \bibinfo {author} {\bibfnamefont {Thomas}\
  \bibnamefont {Hambye}},\ }\bibfield  {title} {\enquote {\bibinfo {title}
  {{Scalar triplet flavored leptogenesis: a systematic approach}},}\ }\href
  {\doibase 10.1088/1475-7516/2014/08/003} {\bibfield  {journal} {\bibinfo
  {journal} {JCAP}\ }\textbf {\bibinfo {volume} {08}},\ \bibinfo {pages} {003}
  (\bibinfo {year} {2014})},\ \Eprint {http://arxiv.org/abs/1401.4347}
  {arXiv:1401.4347 [hep-ph]} \BibitemShut {NoStop}%
\bibitem [{\citenamefont {Lavignac}\ and\ \citenamefont
  {Schmauch}(2015)}]{Lavignac:2015gpa}%
  \BibitemOpen
  \bibfield  {author} {\bibinfo {author} {\bibfnamefont {St\'ephane}\
  \bibnamefont {Lavignac}}\ and\ \bibinfo {author} {\bibfnamefont
  {Beno\^\i{}t}\ \bibnamefont {Schmauch}},\ }\bibfield  {title} {\enquote
  {\bibinfo {title} {{Flavour always matters in scalar triplet
  leptogenesis}},}\ }\href {\doibase 10.1007/JHEP05(2015)124} {\bibfield
  {journal} {\bibinfo  {journal} {JHEP}\ }\textbf {\bibinfo {volume} {05}},\
  \bibinfo {pages} {124} (\bibinfo {year} {2015})},\ \Eprint
  {http://arxiv.org/abs/1503.00629} {arXiv:1503.00629 [hep-ph]} \BibitemShut
  {NoStop}%
\bibitem [{\citenamefont {Aghanim}\ \emph {et~al.}(2020)\citenamefont {Aghanim}
  \emph {et~al.}}]{Planck:2018vyg}%
  \BibitemOpen
  \bibfield  {author} {\bibinfo {author} {\bibfnamefont {N.}~\bibnamefont
  {Aghanim}} \emph {et~al.} (\bibinfo {collaboration} {Planck}),\ }\bibfield
  {title} {\enquote {\bibinfo {title} {{Planck 2018 results. VI. Cosmological
  parameters}},}\ }\href {\doibase 10.1051/0004-6361/201833910} {\bibfield
  {journal} {\bibinfo  {journal} {Astron. Astrophys.}\ }\textbf {\bibinfo
  {volume} {641}},\ \bibinfo {pages} {A6} (\bibinfo {year} {2020})},\ \bibinfo
  {note} {[Erratum: Astron.Astrophys. 652, C4 (2021)]},\ \Eprint
  {http://arxiv.org/abs/1807.06209} {arXiv:1807.06209 [astro-ph.CO]}
  \BibitemShut {NoStop}%
\bibitem [{\citenamefont {Fong}(2022)}]{Fong:2021xmi}%
  \BibitemOpen
  \bibfield  {author} {\bibinfo {author} {\bibfnamefont {Chee~Sheng}\
  \bibnamefont {Fong}},\ }\bibfield  {title} {\enquote {\bibinfo {title}
  {{Cosmic evolution of lepton flavor charges}},}\ }\href {\doibase
  10.1103/PhysRevD.105.043004} {\bibfield  {journal} {\bibinfo  {journal}
  {Phys. Rev. D}\ }\textbf {\bibinfo {volume} {105}},\ \bibinfo {pages}
  {043004} (\bibinfo {year} {2022})},\ \Eprint
  {http://arxiv.org/abs/2109.04478} {arXiv:2109.04478 [hep-ph]} \BibitemShut
  {NoStop}%
\bibitem [{\citenamefont {Garbrecht}\ \emph {et~al.}(2013)\citenamefont
  {Garbrecht}, \citenamefont {Glowna},\ and\ \citenamefont
  {Schwaller}}]{Garbrecht:2013bia}%
  \BibitemOpen
  \bibfield  {author} {\bibinfo {author} {\bibfnamefont {Bj\"orn}\ \bibnamefont
  {Garbrecht}}, \bibinfo {author} {\bibfnamefont {Frank}\ \bibnamefont
  {Glowna}}, \ and\ \bibinfo {author} {\bibfnamefont {Pedro}\ \bibnamefont
  {Schwaller}},\ }\bibfield  {title} {\enquote {\bibinfo {title} {{Scattering
  Rates For Leptogenesis: Damping of Lepton Flavour Coherence and Production of
  Singlet Neutrinos}},}\ }\href {\doibase 10.1016/j.nuclphysb.2013.08.020}
  {\bibfield  {journal} {\bibinfo  {journal} {Nucl. Phys. B}\ }\textbf
  {\bibinfo {volume} {877}},\ \bibinfo {pages} {1--35} (\bibinfo {year}
  {2013})},\ \Eprint {http://arxiv.org/abs/1303.5498} {arXiv:1303.5498
  [hep-ph]} \BibitemShut {NoStop}%
\bibitem [{\citenamefont {Garbrecht}\ and\ \citenamefont
  {Schwaller}(2014)}]{Garbrecht:2014kda}%
  \BibitemOpen
  \bibfield  {author} {\bibinfo {author} {\bibfnamefont {Bj\"orn}\ \bibnamefont
  {Garbrecht}}\ and\ \bibinfo {author} {\bibfnamefont {Pedro}\ \bibnamefont
  {Schwaller}},\ }\bibfield  {title} {\enquote {\bibinfo {title} {{Spectator
  Effects during Leptogenesis in the Strong Washout Regime}},}\ }\href
  {\doibase 10.1088/1475-7516/2014/10/012} {\bibfield  {journal} {\bibinfo
  {journal} {JCAP}\ }\textbf {\bibinfo {volume} {10}},\ \bibinfo {pages} {012}
  (\bibinfo {year} {2014})},\ \Eprint {http://arxiv.org/abs/1404.2915}
  {arXiv:1404.2915 [hep-ph]} \BibitemShut {NoStop}%
\bibitem [{\citenamefont {Hambye}\ \emph {et~al.}(2006)\citenamefont {Hambye},
  \citenamefont {Raidal},\ and\ \citenamefont {Strumia}}]{Hambye:2005tk}%
  \BibitemOpen
  \bibfield  {author} {\bibinfo {author} {\bibfnamefont {Thomas}\ \bibnamefont
  {Hambye}}, \bibinfo {author} {\bibfnamefont {Martti}\ \bibnamefont {Raidal}},
  \ and\ \bibinfo {author} {\bibfnamefont {Alessandro}\ \bibnamefont
  {Strumia}},\ }\bibfield  {title} {\enquote {\bibinfo {title} {{Efficiency and
  maximal CP-asymmetry of scalar triplet leptogenesis}},}\ }\href {\doibase
  10.1016/j.physletb.2005.11.007} {\bibfield  {journal} {\bibinfo  {journal}
  {Phys. Lett. B}\ }\textbf {\bibinfo {volume} {632}},\ \bibinfo {pages}
  {667--674} (\bibinfo {year} {2006})},\ \Eprint
  {http://arxiv.org/abs/hep-ph/0510008} {arXiv:hep-ph/0510008} \BibitemShut
  {NoStop}%
\bibitem [{\citenamefont {D'Onofrio}\ \emph {et~al.}(2014)\citenamefont
  {D'Onofrio}, \citenamefont {Rummukainen},\ and\ \citenamefont
  {Tranberg}}]{DOnofrio:2014rug}%
  \BibitemOpen
  \bibfield  {author} {\bibinfo {author} {\bibfnamefont {Michela}\ \bibnamefont
  {D'Onofrio}}, \bibinfo {author} {\bibfnamefont {Kari}\ \bibnamefont
  {Rummukainen}}, \ and\ \bibinfo {author} {\bibfnamefont {Anders}\
  \bibnamefont {Tranberg}},\ }\bibfield  {title} {\enquote {\bibinfo {title}
  {{Sphaleron Rate in the Minimal Standard Model}},}\ }\href {\doibase
  10.1103/PhysRevLett.113.141602} {\bibfield  {journal} {\bibinfo  {journal}
  {Phys. Rev. Lett.}\ }\textbf {\bibinfo {volume} {113}},\ \bibinfo {pages}
  {141602} (\bibinfo {year} {2014})},\ \Eprint {http://arxiv.org/abs/1404.3565}
  {arXiv:1404.3565 [hep-ph]} \BibitemShut {NoStop}%
\bibitem [{\citenamefont {Fong}(2016)}]{Fong:2015vna}%
  \BibitemOpen
  \bibfield  {author} {\bibinfo {author} {\bibfnamefont {Chee~Sheng}\
  \bibnamefont {Fong}},\ }\bibfield  {title} {\enquote {\bibinfo {title}
  {{Baryogenesis from Symmetry Principle}},}\ }\href {\doibase
  10.1016/j.physletb.2015.11.055} {\bibfield  {journal} {\bibinfo  {journal}
  {Phys. Lett. B}\ }\textbf {\bibinfo {volume} {752}},\ \bibinfo {pages}
  {247--251} (\bibinfo {year} {2016})},\ \Eprint
  {http://arxiv.org/abs/1508.03648} {arXiv:1508.03648 [hep-ph]} \BibitemShut
  {NoStop}%
\bibitem [{\citenamefont {Antusch}\ \emph {et~al.}(2003)\citenamefont
  {Antusch}, \citenamefont {Kersten}, \citenamefont {Lindner},\ and\
  \citenamefont {Ratz}}]{Antusch:2003kp}%
  \BibitemOpen
  \bibfield  {author} {\bibinfo {author} {\bibfnamefont {Stefan}\ \bibnamefont
  {Antusch}}, \bibinfo {author} {\bibfnamefont {J\"orn}\ \bibnamefont
  {Kersten}}, \bibinfo {author} {\bibfnamefont {Manfred}\ \bibnamefont
  {Lindner}}, \ and\ \bibinfo {author} {\bibfnamefont {Michael}\ \bibnamefont
  {Ratz}},\ }\bibfield  {title} {\enquote {\bibinfo {title} {{Running neutrino
  masses, mixings and CP phases: Analytical results and phenomenological
  consequences}},}\ }\href {\doibase 10.1016/j.nuclphysb.2003.09.050}
  {\bibfield  {journal} {\bibinfo  {journal} {Nucl. Phys. B}\ }\textbf
  {\bibinfo {volume} {674}},\ \bibinfo {pages} {401--433} (\bibinfo {year}
  {2003})},\ \Eprint {http://arxiv.org/abs/hep-ph/0305273}
  {arXiv:hep-ph/0305273} \BibitemShut {NoStop}%
\bibitem [{\citenamefont {Chankowski}\ and\ \citenamefont
  {Pokorski}(2002)}]{Chankowski:2001mx}%
  \BibitemOpen
  \bibfield  {author} {\bibinfo {author} {\bibfnamefont {Piotr~H.}\
  \bibnamefont {Chankowski}}\ and\ \bibinfo {author} {\bibfnamefont {Stefan}\
  \bibnamefont {Pokorski}},\ }\bibfield  {title} {\enquote {\bibinfo {title}
  {{Quantum corrections to neutrino masses and mixing angles}},}\ }\href
  {\doibase 10.1142/S0217751X02006109} {\bibfield  {journal} {\bibinfo
  {journal} {Int. J. Mod. Phys. A}\ }\textbf {\bibinfo {volume} {17}},\
  \bibinfo {pages} {575--614} (\bibinfo {year} {2002})},\ \Eprint
  {http://arxiv.org/abs/hep-ph/0110249} {arXiv:hep-ph/0110249} \BibitemShut
  {NoStop}%
\bibitem [{\citenamefont {Ohlsson}\ and\ \citenamefont
  {Zhou}(2014)}]{Ohlsson:2013xva}%
  \BibitemOpen
  \bibfield  {author} {\bibinfo {author} {\bibfnamefont {Tommy}\ \bibnamefont
  {Ohlsson}}\ and\ \bibinfo {author} {\bibfnamefont {Shun}\ \bibnamefont
  {Zhou}},\ }\bibfield  {title} {\enquote {\bibinfo {title} {{Renormalization
  group running of neutrino parameters}},}\ }\href {\doibase
  10.1038/ncomms6153} {\bibfield  {journal} {\bibinfo  {journal} {Nature
  Commun.}\ }\textbf {\bibinfo {volume} {5}},\ \bibinfo {pages} {5153}
  (\bibinfo {year} {2014})},\ \Eprint {http://arxiv.org/abs/1311.3846}
  {arXiv:1311.3846 [hep-ph]} \BibitemShut {NoStop}%
\bibitem [{\citenamefont {Esteban}\ \emph {et~al.}(2024)\citenamefont
  {Esteban}, \citenamefont {Gonzalez-Garcia}, \citenamefont {Maltoni},
  \citenamefont {Martinez-Soler}, \citenamefont {Pinheiro},\ and\ \citenamefont
  {Schwetz}}]{Esteban:2024eli}%
  \BibitemOpen
  \bibfield  {author} {\bibinfo {author} {\bibfnamefont {Ivan}\ \bibnamefont
  {Esteban}}, \bibinfo {author} {\bibfnamefont {M.~C.}\ \bibnamefont
  {Gonzalez-Garcia}}, \bibinfo {author} {\bibfnamefont {Michele}\ \bibnamefont
  {Maltoni}}, \bibinfo {author} {\bibfnamefont {Ivan}\ \bibnamefont
  {Martinez-Soler}}, \bibinfo {author} {\bibfnamefont {Jo\~ao~Paulo}\
  \bibnamefont {Pinheiro}}, \ and\ \bibinfo {author} {\bibfnamefont {Thomas}\
  \bibnamefont {Schwetz}},\ }\bibfield  {title} {\enquote {\bibinfo {title}
  {{NuFit-6.0: updated global analysis of three-flavor neutrino
  oscillations}},}\ }\href {\doibase 10.1007/JHEP12(2024)216} {\bibfield
  {journal} {\bibinfo  {journal} {JHEP}\ }\textbf {\bibinfo {volume} {12}},\
  \bibinfo {pages} {216} (\bibinfo {year} {2024})},\ \Eprint
  {http://arxiv.org/abs/2410.05380} {arXiv:2410.05380 [hep-ph]} \BibitemShut
  {NoStop}%
\bibitem [{\citenamefont {Feruglio}\ \emph {et~al.}(2015)\citenamefont
  {Feruglio}, \citenamefont {Patel},\ and\ \citenamefont
  {Vicino}}]{Feruglio:2015iua}%
  \BibitemOpen
  \bibfield  {author} {\bibinfo {author} {\bibfnamefont {Ferruccio}\
  \bibnamefont {Feruglio}}, \bibinfo {author} {\bibfnamefont {Ketan~M.}\
  \bibnamefont {Patel}}, \ and\ \bibinfo {author} {\bibfnamefont {Denise}\
  \bibnamefont {Vicino}},\ }\bibfield  {title} {\enquote {\bibinfo {title} {{A
  realistic pattern of fermion masses from a five-dimensional SO(10) model}},}\
  }\href {\doibase 10.1007/JHEP09(2015)040} {\bibfield  {journal} {\bibinfo
  {journal} {JHEP}\ }\textbf {\bibinfo {volume} {09}},\ \bibinfo {pages} {040}
  (\bibinfo {year} {2015})},\ \Eprint {http://arxiv.org/abs/1507.00669}
  {arXiv:1507.00669 [hep-ph]} \BibitemShut {NoStop}%
\bibitem [{\citenamefont {Babu}\ \emph {et~al.}(2018)\citenamefont {Babu},
  \citenamefont {Bajc},\ and\ \citenamefont {Saad}}]{Babu:2018tfi}%
  \BibitemOpen
  \bibfield  {author} {\bibinfo {author} {\bibfnamefont {K.~S.}\ \bibnamefont
  {Babu}}, \bibinfo {author} {\bibfnamefont {Borut}\ \bibnamefont {Bajc}}, \
  and\ \bibinfo {author} {\bibfnamefont {Shaikh}\ \bibnamefont {Saad}},\
  }\bibfield  {title} {\enquote {\bibinfo {title} {{Resurrecting Minimal Yukawa
  Sector of SUSY SO(10)}},}\ }\href {\doibase 10.1007/JHEP10(2018)135}
  {\bibfield  {journal} {\bibinfo  {journal} {JHEP}\ }\textbf {\bibinfo
  {volume} {10}},\ \bibinfo {pages} {135} (\bibinfo {year} {2018})},\ \Eprint
  {http://arxiv.org/abs/1805.10631} {arXiv:1805.10631 [hep-ph]} \BibitemShut
  {NoStop}%
\bibitem [{\citenamefont {Abe}\ \emph {et~al.}(2023)\citenamefont {Abe} \emph
  {et~al.}}]{KamLAND-Zen:2022tow}%
  \BibitemOpen
  \bibfield  {author} {\bibinfo {author} {\bibfnamefont {S.}~\bibnamefont
  {Abe}} \emph {et~al.} (\bibinfo {collaboration} {KamLAND-Zen}),\ }\bibfield
  {title} {\enquote {\bibinfo {title} {{Search for the Majorana Nature of
  Neutrinos in the Inverted Mass Ordering Region with KamLAND-Zen}},}\ }\href
  {\doibase 10.1103/PhysRevLett.130.051801} {\bibfield  {journal} {\bibinfo
  {journal} {Phys. Rev. Lett.}\ }\textbf {\bibinfo {volume} {130}},\ \bibinfo
  {pages} {051801} (\bibinfo {year} {2023})},\ \Eprint
  {http://arxiv.org/abs/2203.02139} {arXiv:2203.02139 [hep-ex]} \BibitemShut
  {NoStop}%
\bibitem [{\citenamefont {Shao}\ \emph {et~al.}(2025)\citenamefont {Shao},
  \citenamefont {Givans}, \citenamefont {Dunkley}, \citenamefont
  {Madhavacheril}, \citenamefont {Qu}, \citenamefont {Farren},\ and\
  \citenamefont {Sherwin}}]{Shao:2024mag}%
  \BibitemOpen
  \bibfield  {author} {\bibinfo {author} {\bibfnamefont {Helen}\ \bibnamefont
  {Shao}}, \bibinfo {author} {\bibfnamefont {Jahmour~J.}\ \bibnamefont
  {Givans}}, \bibinfo {author} {\bibfnamefont {Jo}~\bibnamefont {Dunkley}},
  \bibinfo {author} {\bibfnamefont {Mathew}\ \bibnamefont {Madhavacheril}},
  \bibinfo {author} {\bibfnamefont {Frank~J.}\ \bibnamefont {Qu}}, \bibinfo
  {author} {\bibfnamefont {Gerrit}\ \bibnamefont {Farren}}, \ and\ \bibinfo
  {author} {\bibfnamefont {Blake}\ \bibnamefont {Sherwin}},\ }\bibfield
  {title} {\enquote {\bibinfo {title} {{Cosmological limits on the neutrino
  mass sum for beyond-\ensuremath{\Lambda}CDM models}},}\ }\href {\doibase
  10.1103/PhysRevD.111.083535} {\bibfield  {journal} {\bibinfo  {journal}
  {Phys. Rev. D}\ }\textbf {\bibinfo {volume} {111}},\ \bibinfo {pages}
  {083535} (\bibinfo {year} {2025})},\ \Eprint
  {http://arxiv.org/abs/2409.02295} {arXiv:2409.02295 [astro-ph.CO]}
  \BibitemShut {NoStop}%
\end{thebibliography}%

\end{document}